\documentclass[aps,prl,amsmath,amssymb,twocolumn]{revtex4-2}
\usepackage[english]{babel}
\usepackage[utf8]{inputenc}
\usepackage{amsfonts}
\usepackage[T1]{fontenc}
\usepackage[pdftex]{graphicx}
\usepackage{fancyhdr}
\usepackage{hyperref}
\usepackage{url}
\usepackage[titletoc,title]{appendix}
\usepackage{epstopdf}
\usepackage{amsmath}
\usepackage{amssymb}
\usepackage{dsfont}
\usepackage{mathtools}
\usepackage{bbold}
\usepackage{xcolor}
\usepackage{tikz}
\usepackage{cancel}
\usetikzlibrary{arrows.meta}

\hypersetup{
    bookmarks=false,         
    unicode=false,          
    pdftoolbar=false,        
    pdfmenubar=true,        
    pdffitwindow=false,     
    pdfstartview={FitH},    
    pdftitle={Pairing-induced phase transition in the non-reciprocal Kitaev chain},    
    pdfauthor={P. Brighi and A. Nunnenkamp},     
    pdfsubject={},   
    pdfcreator={},   
    pdfproducer={}, 
    pdfkeywords={},
    pdfnewwindow=true,      
    colorlinks=true,       
    linkcolor=black,          
    citecolor=blue,        
    filecolor=magenta,      
    urlcolor=blue           
}

\graphicspath{{./Figures/}}

\newcommand{\id}{\text{\usefont{U}{bbold}{m}{n}1}}
\newcommand{\ket}[1]{|#1\rangle}
\newcommand{\bra}[1]{\langle#1|}

\newcommand{\C}[1]{\hat{c}_{#1}}

\newcommand{\Cdag}[1]{\hat{c}^\dagger_{#1}}

\makeatletter
\newsavebox{\@brx}
\newcommand{\llangle}[1][]{\savebox{\@brx}{\(\m@th{#1\langle}\)}%
  \mathopen{\copy\@brx\kern-0.5\wd\@brx\usebox{\@brx}}}
\newcommand{\rrangle}[1][]{\savebox{\@brx}{\(\m@th{#1\rangle}\)}%
  \mathclose{\copy\@brx\kern-0.5\wd\@brx\usebox{\@brx}}}
\makeatother

\newcommand{\superket}[1]{\Vert {#1} \rrangle}

\newcommand{\mytitle}{Pairing-induced phase transition in the non-reciprocal Kitaev chain}

\begin{document}

\title{\mytitle}
\author{Pietro Brighi}
\affiliation{Faculty of Physics, University of Vienna, Boltzmanngasse 5, 1090 Vienna, Austria}
\author{Andreas Nunnenkamp}
\affiliation{Faculty of Physics, University of Vienna, Boltzmanngasse 5, 1090 Vienna, Austria}
\date{\today}

\begin{abstract}
    Investigating the robustness of non-reciprocity in the presence of competing interactions is central to understanding non-reciprocal quantum matter.
    In this work, we use reservoir engineering to induce non-reciprocal hopping and pairing in the fermionic Kitaev chain, and reveal the emergence of a pairing-induced phase transition.
    The two phases appear in the spectrum of the non-Hermitian Kitaev Hamiltonian describing the dynamics of correlations, separated by an exceptional point.
    In the non-reciprocal phase, dynamics are characterized by directionality and slow relaxation, and the steady state supports non-reciprocal density and spatial correlations.
    At strong pairing, we uncover an unexpected density wave phase, featuring short relaxation times, a modulation in particle occupation and strikingly different correlation spreading depending on pairing non-reciprocity.
    Our work highlights the non-trivial breakdown of non-reciprocity due to superconducting pairing and invites experimental investigation of non-reciprocal fermionic systems.
\end{abstract}

\maketitle

\textit{Introduction ---}
Systems where the degrees of freedom affect one another in a non-reciprocal way give rise to fascinating phases and phase transitions~\cite{Vitelli2021,Vitelli2025,Clerk2025,Brunelli2025}, interesting dynamical phenomena~\cite{Bergholtz2022,Brunelli2025,Hanai2024,Brighi2024} and amplification~\cite{Porras2019,Wanjura2020}.
In the context of open quantum systems, reservoir engineering~\cite{Zoller1996} has been proposed as a practical route to non-reciprocity~\cite{Clerk2015}.
In particular, non-reciprocal hopping has received considerable attention~\cite{Hatano1996,Porras2019,Wanjura2020,Clerk2022,Clerk2023}, due to the presence of the non-Hermitian skin effect~\cite{Lee2016,Wang2018,Chen2022} and of non-Hermitian topology~\cite{Ueda2018,Ueda2019a,Ryu2022,Nunnenkamp2023,Bergholtz2021}.

In spite of these efforts, a deeper understanding of non-reciprocal quantum matter calls for the investigation of minimal models, where competing terms challenge non-reciprocal hopping.
In this context, the role of particle-particle interactions has been explored both within the no-click limit of non-Hermitian Hamiltonians~\cite{Zhu2020,Neupert2022,Brzezinska2022,Pan2023,Longhi2023,Chen2024,Bergholtz2024,Lee2024,Li2024} and at the Lindblad master equation level~\cite{Hanai2024,Brighi2024,Brunelli2025}; these systems, however, often lack analytically tractable solutions, and one has to resort to numerical methods, which are limited in system size.

Here, instead, we consider the quadratic superconducting pairing interaction as the process competing with non-reciprocal hopping.
Using reservoir engineering, we introduce a version of the celebrated Kitaev chain~\cite{Kitaev2003} where both hopping and pairing can be made non-reciprocal.
This is qualitatively different from the behavior of the bosonic Kitaev chain, which supports non-reciprocity even in absence of reservoir engineering~\cite{Clerk2018,Nunnenkamp2023BKC}, as recently tested experimentally~\cite{Verhagen2024,Wilson2024}.
Our reservoir engineering approach overcomes the limitations of the no-click limit used in recent studies of the non-Hermitian Kitaev chain, where non-reciprocal hopping or pairing~\cite{Ezawa2019,Liu2021,Wang2021,Black-Schaffer2023}, imaginary chemical potential~\cite{Tong2015,Lu2016,Yuce2016,Wunner2017,Ueda2018b,Ueda2019a,Katsura2019} or interactions~\cite{Lado2023} have shown the presence of various topological phases and edge modes enhancement.  
Our work explores a direction different from that of previous reservoir engineering schemes for the Kitaev chain, mostly focused on steady state preparation~\cite{Zoller2011}, or on stability of the Majorana modes to coupling with the environment~\cite{Zoller2015,Divincenzo2015}.

\begin{figure}[b]
    \centering
    \includegraphics[width=0.99\linewidth]{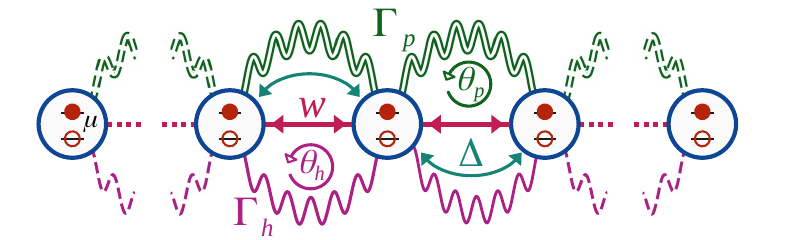}
    \caption{\label{Fig:model cartoon}
    Schematic depiction of the model.
    Each site can host at most one spinless fermion, with an energy cost corresponding to the chemical potential $\mu$.
    Neighboring sites are coupled coherently through a hopping term $w$ and a pairing interaction $\Delta$ which breaks $U(1)$ symmetry [Eq.~(\ref{Eq:H})].
    The dissipative part of the model gives rise to incoherent hopping and pairing terms with rates $\Gamma_h$ and $\Gamma_p$, and phases $\theta_h$ and $\theta_p$, respectively [Eq.~(\ref{Eq:L})].
    These interfere with the analogous coherent processes, generating non-reciprocity.
    }
\end{figure}

Studying the dynamics and steady state under open boundary conditions, we identify a pairing-induced phase transition.
At weak pairing, the system exhibits a non-reciprocal phase characterized by directional dynamics, slow relaxation, and steady-state non-reciprocity.
At the critical point, a density-wave phase emerges from the breakdown of non-reciprocity, featuring a modulation of particle density both in dynamics and steady state, as well as short relaxation times.
This phase further hosts two different regimes, depending on pairing non-reciprocity.
These findings show that introducing pairing interactions in non-reciprocal hopping models qualitatively affects the many-body dynamics and phase transitions, remaining in the domain of quadratic, solvable, problems.

\textit{Model ---}
We consider a one-dimensional chain of spinless fermions whose dynamics are governed by the fermionic Kitaev chain~\cite{Kitaev2003} 
\begin{equation}
    \label{Eq:H}
    \hat{H} \!=\! \sum_j\!  -w\!\left(\Cdag{j}\C{j+1}\! + \! \text{H.c.}\!\right)\! +\! \left(\!\Delta\Cdag{j}\Cdag{j+1} \!+\! \text{H.c} \right) \!-\!\mu\Cdag{j}\C{j},
\end{equation}
where $w$ and $\Delta$ are the hopping and pairing amplitude, respectively, and $\mu$ is the chemical potential.
The Hamiltonian has particle-hole symmetry and describes a $p$-wave superconductor which hosts a topological phase, characterized by the presence of two Majorana edge modes, which have been proposed as potential platform for quantum computing~\cite{DasSarma2008}.

To understand the consequences of superconducting pairing on the non-reciprocal behavior of free fermions~\cite{Clerk2022,Brunelli2025} we couple the system to an engineered environment, described by two jump operators leading to non-reciprocal hopping and non-reciprocal pairing
\begin{equation}
    \label{Eq:L}
    \hat{L}^{(h)}_j \!=\! \sqrt{\Gamma_h}\!\left(\C{j}\! +\! e^{\imath\theta_h}\C{j+1}\right), \;
    \hat{L}^{(p)}_j\! =\! \sqrt{\Gamma_p}\!\left(\C{j}\! +\! e^{\imath\theta_p}\Cdag{j+1}\right).
\end{equation}
These operators describe incoherent processes which couple sites $j$ and $j+1$ with rate $\Gamma_h$ and $\Gamma_p$.
Depending on the phases $\theta_h$ and $\theta_p$, these can interfere with coherent processes and generate non-reciprocity in the hopping~\cite{Clerk2015,Porras2019,Wanjura2020} and pairing~\cite{Clerk2023a,Metelmann2023}, respectively.
In Figure~\ref{Fig:model cartoon} we pictorially show the interplay of the various terms defining the model.

\textit{Non-Hermitian Kitaev chain ---}
Since the Hamiltonian~(\ref{Eq:H}) is quadratic and the dissipative terms~(\ref{Eq:L}) are linear in fermionic operators, all relevant information about dynamics and steady state is encoded in the correlation functions.
Due to superconducting pairing, the equations of motion close upon introducing the anomalous correlator $\langle\Cdag{\ell}\Cdag{m}\rangle(t)$.

We combine all correlation functions into a correlation matrix $\mathcal{C}$, and study its equations of motion using the Lindblad master equation for operators $
    \frac{d\langle\hat{O}\rangle}{dt} = \imath\langle[\hat{H},\hat{O}]\rangle + \sum_{j,n} \langle\hat{L}_j^{(n)\dagger}\hat{O}\hat{L}^{(n)}_j\rangle - \frac{1}{2}\langle\{\hat{L}^{(n)\dagger}_j\hat{L}^{(n)}_j,\hat{O}\}\rangle $
to obtain a von-Neumann-like equation for its dynamics $
    \dot{\mathcal{C}} = -\imath \left(\mathbb{H} \mathcal{C} - \mathcal{C}\mathbb{H}^\dagger\right) + \mathbb{F}$~\cite{Supplementary}.
In OBC, the non-Hermitian dynamical matrix $\mathbb{H}$ and the noise term $\mathbb{F}$ are $2N \times 2N$ matrices.
In particular,
\begin{equation}
    \mathbb{H}_{mn} \!=\! 
    \begin{cases}
        \begin{split} &\!w_R\delta_{m n-1} \!+\! w_L\delta_{m n+1} \!-\! \Delta^*_R\delta_{m n-N-1} \\ &\!+\! \Delta^*_L\delta_{m n-N +1}\! -\! [\imath(\Gamma_h \!+\! \Gamma_p) \!- \!\mu]\delta_{mn}
        \end{split}
        & \mkern-18mu m\leq N \\
        \\
        \begin{split}&\!-\!w^*_R\delta_{m n-1} \!-\! w^*_L\delta_{m n+1} \!+\! \Delta_R\delta_{m n+N-1} \\ &\!-\! \Delta_L\delta_{m n+N+1}\! -\! [\imath(\Gamma_h \!+\! \Gamma_p) \!+\! \mu] \delta_{mn} 
        \end{split}
        & \mkern-18mu m > N
    \end{cases},
\end{equation}
which corresponds to a non-Hermitian Kitaev chain where both hopping and pairing can be non-reciprocal, $w_{R/L} = w - \imath\frac{\Gamma_h}{2}e^{\mp\imath\theta_h}$ and $\Delta_{R/L} = \Delta \mp \imath\frac{\Gamma_p}{2}e^{\imath\theta_p}$.
Since the pairing term breaks $U(1)$ symmetry, the system has a non-trivial steady state ($\mathbb{F}\neq \mathbb{0}$) if $\Delta \neq 0$ or $\Gamma_p \neq 0$~\cite{Supplementary}.
In PBC the dynamical matrix and the noise term are $2\times 2$ momentum dependent matrices (see End Matter).
Thus, through the jump operators~(\ref{Eq:L}) the non-Hermitian Kitaev chain naturally emerges as the dynamical matrix describing the time evolution of correlations, overcoming the postselection problem of the no-click limit.

\begin{figure}[t]
    \centering
    \includegraphics[width=0.99\linewidth]{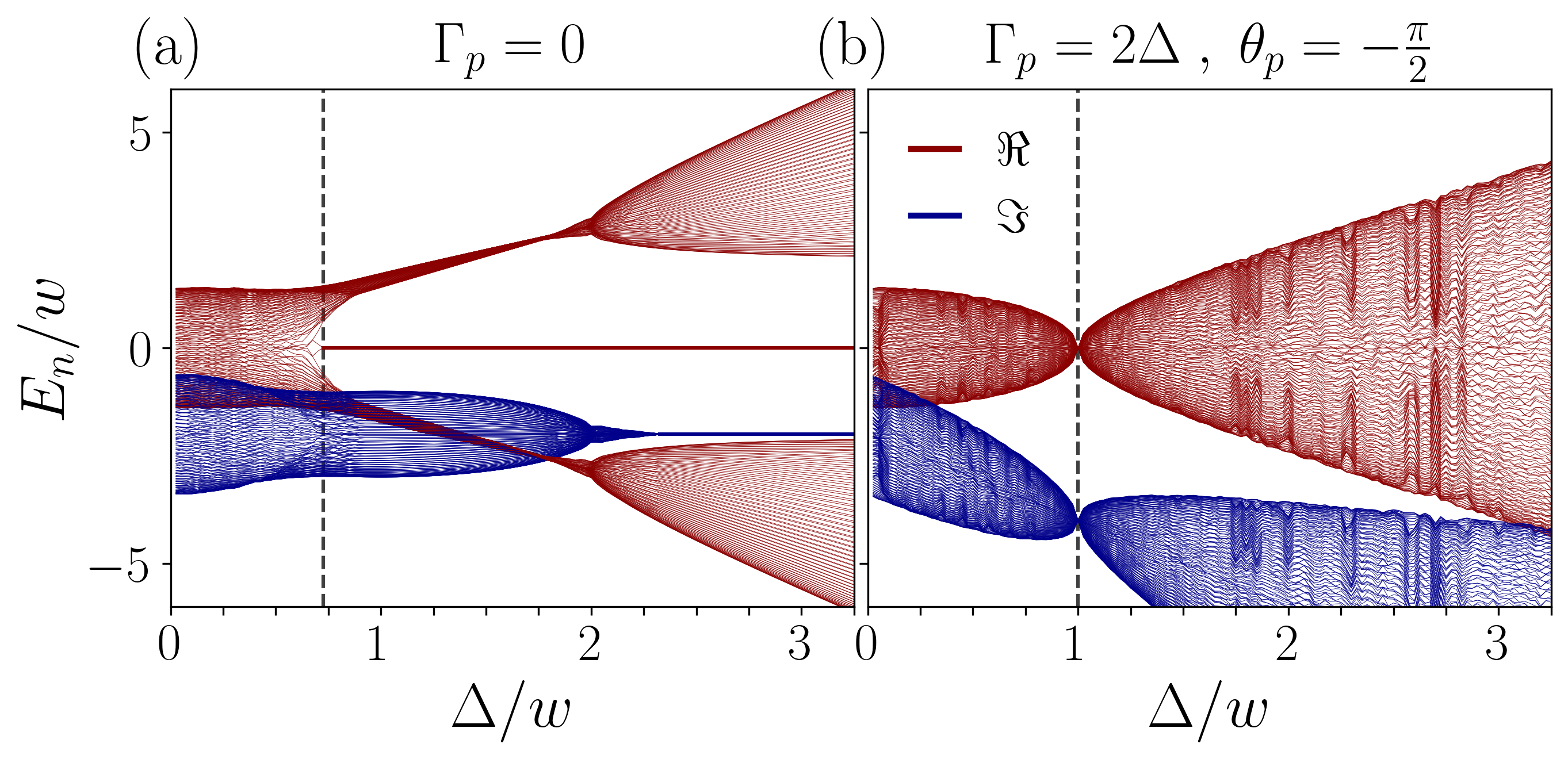}
    \caption{\label{Fig:spectra} 
    Real and imaginary part of the OBC spectrum, red and blue curves, for coherent (a) and non-reciprocal pairing (b) as a function of $\Delta$ for $\Gamma_h = 2w$, $\theta_h = \frac{\pi}{2}$ and $\mu = 0$; system size is $N = 100$.
    (a) For coherent pairing the spectrum shows a transition, marked by a gap opening in the real part.
    The gapped phase hosts two eigenvalues with $\Re[E_n]=0$, reminiscent of the Majorana zero modes.
    (b) For non-reciprocal pairing the complex spectrum is gapless.
    However, an exceptional point appears at $\Delta = w$.
    }
\end{figure}

In the following, we focus on the effect of pairing on a system with purely non-reciprocal hopping  $\Gamma_h = 2w$, $\theta_h = \frac{\pi}{2}$ and $\mu = 0$, discussing the case of purely coherent pairing, $\Gamma_p = 0$, and non-reciprocal pairing with the same directionality as the hopping, $\Gamma_p = 2\Delta,\;\theta_p = -\frac{\pi}{2}$.
These two cases simplify the study of the phase diagram, and showcase the effect of coherent and non-reciprocal pairing.

\begin{figure*}
    \centering
    \includegraphics[width=0.99\textwidth]{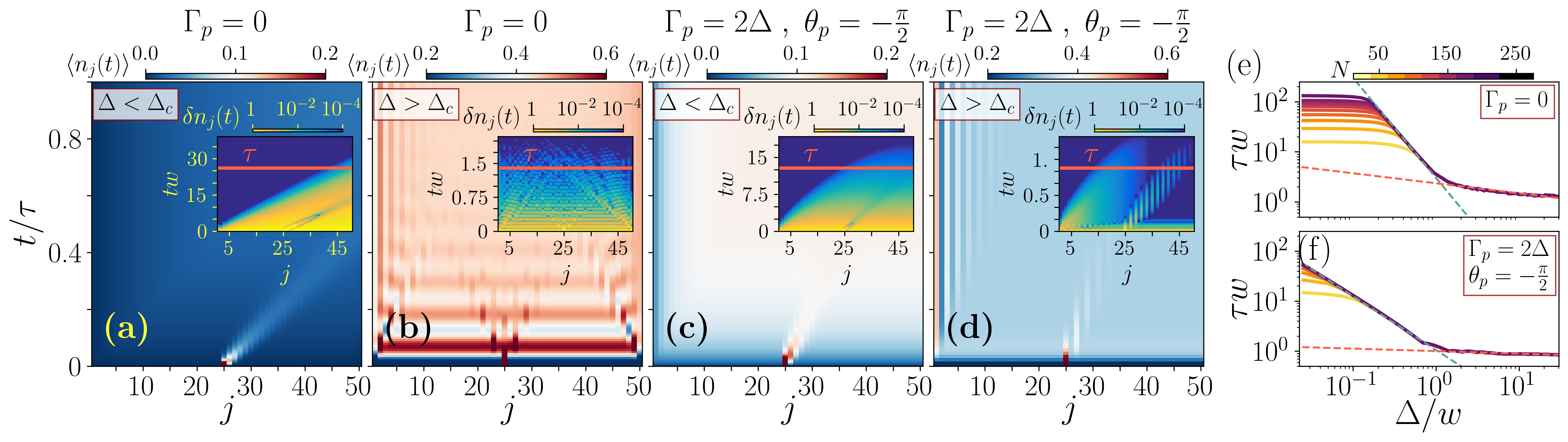}
    \caption{\label{Fig:nt OBC Delta}
    (a)-(d) Density dynamics in the non-reciprocal Kitaev chain.
    As $\Delta <\Delta_c = w$ ($\Delta = 0.1w$ is shown here) both the purely coherent ($\Gamma_p = 0$) and non-reciprocal ($\Gamma_p = 2\Delta$) pairing cases show a clear unidirectional lightcone.
    Non-reciprocity is further highlighted by the relaxation to the steady state shown in the insets.
    At $\Delta>\Delta_c$ ($\Delta = 10w$ is shown here) a density wave pattern emerges.
    Crucially, when $\Gamma_p = 2\Delta$ the density wave is completely non-reciprocal and spreads to the right only.
    (e)-(f) Relaxation time $\tau$ as a function of $\Delta$ for different system sizes $N$.
    Both for $\Gamma_p = 0$ (e) and for non-reciprocal pairing (f) the relaxation time decays as a fast power-law in the non-reciprocal phase $\Delta<w$ (light blue dashed lines).
    In the density wave phase, the power-law drastically changes exponent, showing a very slow decay (red dashed lines).
    }
\end{figure*}

In Figure~\ref{Fig:spectra} we show the real and imaginary part of the OBC spectrum of $\mathbb{H}$ (red and blue curves) as a function of $\Delta$.
For purely coherent pairing~(a), the spectrum presents a transition at $\Delta_c\to w$ as $N\to \infty$~\cite{Supplementary} corresponding to a gap opening in its real part.
Interestingly, in the gapped phase the real part of the spectrum hosts zero modes, reminiscent of the topologically protected Majorana modes in the Kitaev chain.
At larger $\Delta \gtrsim 2w$ the imaginary part of the spectrum becomes single-valued $\Im[E_n] = -\Gamma_h$, this feature however emerges at larger $\Delta$ as $N$ is increased, suggesting its absence in the thermodynamic limit~\cite{Supplementary}.
Surprisingly, introducing non-reciprocal pairing~(b), the spectrum presents an $N$-fold exceptional point at $\Delta = w$, where $E_n = -2\imath(w+\Delta)$, which separates two gapless phases.

\textit{Vectorization of $\mathcal{C}$ ---}
To solve the dynamics and find the steady state, we need to integrate the equation of motion for the correlation matrix.
It is then convenient to \textit{vectorize} the correlation matrix $\superket{\mathcal{C}} =\text{vec}(\mathcal{C}),\; \text{vec}(\mathcal{C})_{m + 2N\times n} = \mathcal{C}_{mn};$ in this convention, the superoperator determining dynamics and steady state becomes a matrix $\mathcal{H} = \id\otimes\mathbb{H} - \mathbb{H}^*\otimes\id.$
We can then formally solve the equations of motion
\begin{equation}
    \label{Eq:vec(C)(t) OBC}
    \superket{\mathcal{C}(t)} = e^{-\imath\mathcal{H}t}\superket{\mathcal{C}_0} + \imath \mathcal{H}^{-1}\left(e^{-\imath\mathcal{H}t} - 1\right) \superket{\mathbb{F}} 
\end{equation}
and find the steady state
\begin{equation}
    \label{Eq:C ss OBC}
    \superket{\mathcal{C}_{ss}} = -\imath\mathcal{H}^{-1}\superket{\mathbb{F}}.
\end{equation}

The spectrum of $\mathcal{H}$, determining the dynamics, can be obtained from the spectrum of $\mathbb{H}$.
Let $\ket{r_n}$ be a right eigenvector of $\mathbb{H}$, and let $\bra{\ell_m}$ be a left eigenvector of $\mathbb{H}^\dagger$, with eigenvalues $E_n$ and $E^*_m$, respectively.
The vectorized rank-one matrix $\superket{R_{mn}} = \text{vec}(\ket{r_n}\bra{\ell_m})$, then, is a right eigenvector of $\mathcal{H}$ with eigenvalue $\mathcal{E}_{nm} = E_n - E^*_m$
\begin{equation}
    \label{Eq:cal H eigenvalues}
    \mathcal{H}\superket{R_{mn}} = \mathbb{H}\ket{r_n}\bra{\ell_m} - \ket{r_n}\bra{\ell_m}\mathbb{H}^\dagger = \mathcal{E}_{mn}\superket{R_{nm}}.
\end{equation}

\begin{figure*}
    \centering
    \includegraphics[width=.99\textwidth]{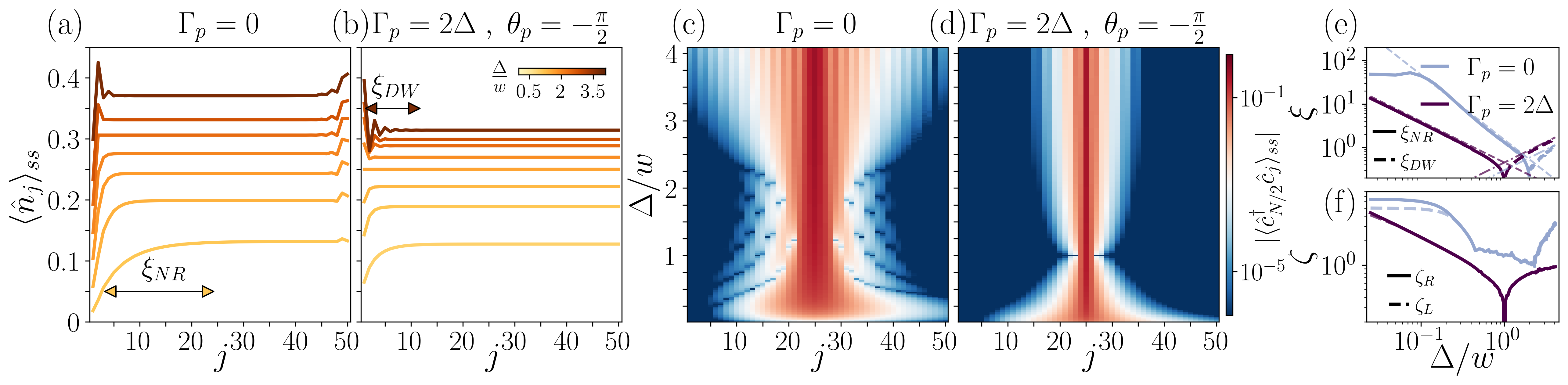}
    \caption{\label{Fig:OBC steady state}
    (a)-(b) Steady state density as a function of pairing amplitude $\Delta$.
    In the non-reciprocal phase $\langle\hat{n}_j\rangle_{ss}$ is inhomogeneous up to a characteristic lengthscale $\xi_{NR}$, while in the density wave phase it acquires a modulation penetrating from the boundaries up to a lengthscale $\xi_{DW}$.
    (c)-(d) Central site correlation functions decay asymmetrically in the non-reciprocal phase and depend strongly on $\Gamma_p$ in the density wave phase.
    (e) Characteristic lenghtscales $\xi_{NR}$ and $\xi_{DW}$ have opposite behavior across the transition: $\xi_{NR}$ decays as a power-law and vanishes at the critical point, while $\xi_{DW}$ increases linearly in the density wave phase.
    (f) Correlation functions decay length $\zeta$. 
    Deep in the non-reciprocal phase, decay is asymmetric, $\zeta_R\neq \zeta_L$, while in the density wave phase $\zeta$ increases for $\Gamma_p =0$ and it tends to saturate for non-reciprocal pairing.
    }
\end{figure*}

\textit{Pairing-induced phase transition ---}
From the transition in the spectrum of $\mathbb{H}$, then, we expect to observe a qualitative transition in the dynamics of the system, originating from the competition between non-reciprocal hopping and pairing.
On one hand, non-reciprocal hopping induces directional particle motion~\cite{Brighi2024} and their accumulation on one side of the chain~\cite{Clerk2022,Brunelli2025}.
On the other, pairing injects pairs of particles everywhere in the chain, thus conflicting with non-reciprocity in the steady state, similar to the effect of incoherent gain discussed in Ref.~\cite{Clerk2022}.
In the following, we will show results for the non-reciprocal Kitaev chain in OBC, where the pairing-induced transition is sharpest.
The interested reader can find details about the transition in PBC in the End Matter and Supplemental Material~\cite{Supplementary}.


To observe the transition, we consider a simple pure initial state $\rho_0 = \ket{\psi_0}\bra{\psi_0},\;\ket{\psi_0} = \Cdag{N/2}\ket{\text{Vac}}$, where a single fermion is initialized in the middle of the chain, and investigate its density dynamics $\langle\hat{n}_j(t)\rangle$ obtained from Eq.~(\ref{Eq:vec(C)(t) OBC}).

In the weak pairing limit shown in Figure~\ref{Fig:nt OBC Delta}~(a),(c) ($\Delta = 0.1w$) particles spread along a unidirectional lightcone, indicating non-reciprocal dynamics for $\Delta<\Delta_c = w$.
In the pairing-dominated regime at large $\Delta=10w$ short time dynamics reveal a qualitatively different behavior characterized by the emergence of a density wave pattern (b),(d).
The density modulation originates and spreads ballistically from defects breaking translational invariance, i.e.~the boundaries and the particle at $j=\frac{N}{2}$.
While for $\Gamma_p = 0$ the density wave is reciprocal, for non-reciprocal pairing it spreads only along the right branch of the defect wavefront, resulting in a non-reciprocal density wave pattern.

The phase transition dramatically affects also the approach to the steady state, shown in the insets of Figure~\ref{Fig:nt OBC Delta} via $\delta n_j(t) = |\langle\hat{n}_j(t)\rangle - \langle\hat{n}_j\rangle_{ss}|/\langle\hat{n}_j\rangle_{ss}$.
To capture the relaxation behavior, we introduce the relaxation time $\tau$: $\max_j[\delta n_j(\tau)]<\epsilon$, with $\epsilon=10^{-3}$.
In Figure~\ref{Fig:nt OBC Delta}(e),(f) we show the relaxation time as a function of the pairing amplitude $\Delta$ and for different system sizes $N$.
Both in the coherent pairing [$\Gamma_p = 0$, panel (e)] and in the non-reciprocal pairing [$\Gamma_p = 2\Delta$, panel (f)] regimes, the relaxation time presents a sharp transition at $\Delta_c\approx w$.
In the non-reciprocal phase the relaxation time is large and diverges as $\Delta\to 0$ as a power law $\Delta^{-\alpha(\Gamma_p)}$ with $\alpha = 2$ for $\Gamma_p = 0$ and $\alpha = 1$ for $\Gamma_p = 2\Delta$ (light blue dashed lines).
Past the critical point, the relaxation time abruptly changes its power law decay to a much slower decay as $\Delta\to\infty$ (red dashed lines) \footnote{Checking the decay from random initial states this behavior remains qualitatively the same.}. 

Performing system size scaling, we observe that in the non-reciprocal phase $\tau$ follows the $\Delta^{-\alpha}$ behavior only up to a certain pairing amplitude vanishing with $N$.
This suggests the emergence of a pairing-dependent lengthscale $\xi_\Delta$ governing the relaxation and diverging as $\Delta\to 0$.
As long as $N>\xi_\Delta$ the relaxation time follows the expected power-law behavior, while as $\Delta$ is decreased and $\xi_\Delta\approx N$ the relaxation time saturates to a system size dependent value.
This behavior is more evident in the coherent pairing case, suggesting that $\xi_\Delta$ is due to non-reciprocal hopping, and indicating that non-reciprocal pairing reduces this lengthscale.

We now focus on the transition in the steady state~(\ref{Eq:C ss OBC}).
Since the steady state is determined by the inverse of the dynamical matrix, Eq.~(\ref{Eq:C ss OBC}), its relation with the spectrum of $\mathcal{H}$ is less straightforward~\cite{Nunnenkamp2023,Balducci2025}.
Nevertheless, the pairing-induced breakdown of non-reciprocity is directly observable, as we show in Figure~\ref{Fig:OBC steady state} focusing on the particle density [panels (a)-(b)] and on the correlation function from the central site $\langle\Cdag{N/2}\C{j}\rangle_{ss}$ [panels (c)-(d)].


In the non-reciprocal phase, the steady state shows inhomogeneous particle density featuring a less populated region close to the left boundary.
We characterize non-reciprocity through the lengthscale $\xi_{NR}$ capturing the penetration of this region into the bulk.
As we show as a solid line in panel (e), $\xi_{NR}$ decays as a power-law up to $\Delta = 2w$ for $\Gamma_p = 0$ and $\Delta = w$ for $\Gamma_p = 2\Delta$.
At these critical points the density profile becomes flat, and $\xi_{NR}\to0$.
In the End Matter we report system size scaling of $\xi_{NR}$, showing that for $\Gamma_p = 0$ it deviates from the power-law behavior once it becomes comparable with system size at $\Delta\ll w$.
We identify this lengthscale with $\xi_\Delta$ yielding finite size effects in the behavior of the relaxation time described above.


As we show in Figure~\ref{Fig:OBC steady state}~(c)-(d), the non-reciprocal phase is characterized by asymmetric decay of one-particle correlations, especially evident for $\Gamma_p = 0$.
In panel~(f), we compare the exponential decay of the correlation functions $|\langle\Cdag{N/2}\C{j}\rangle_{ss}|\propto e^{-\frac{|j-N/2|}{\zeta}}$ on the left ($\zeta_L$) and right ($\zeta_R$) halves of the chain as a solid and dashed line, respectively.
Deep in the non-reciprocal phase, the two lengthscales are different, highlighting the slower decay of correlations in the right half of the chain.
Notably, this non-reciprocity fades out before the phase boundaries are reached.


For $\Gamma_p = 0$ a critical region emerges at $w<\Delta<2w$, where the non-reciprocity in the density and correlations is very weak, but the features of the pairing-dominated phase are not yet fully developed.
On the other hand, for non-reciprocal pairing the transition is sharp, and at the critical point $\Delta = w$ the density is completely flat $\langle\hat{n}_j\rangle_{ss} = 1/4$ and the correlations become extremely short range, with $\langle\Cdag{N/2}\C{j}\rangle_{ss} = 0$ for $j\lessgtr N/2 \mp 2$.


At large $\Delta\gg w$, the steady state presents a density wave order, akin to the one observed in dynamics, penetrating in the bulk from the boundaries up to a characteristic lengthscale $\xi_{DW}$.
This feature is due to the pairing interaction and is completely absent in other scenarios where non-reciprocity is destroyed e.g.~by incoherent gain~\cite{Clerk2022,Brunelli2025}.
Coherent pairing leads to the emergence of density modulation on both ends of the chain, while purely non-reciprocal pairing results in density wave order only close to the left boundary.
As we report in panel (e) $\xi_{DW}$ grows linearly with $\Delta$ and has small finite size effects, thus suggesting that at large $\Delta$ a considerable fraction of the chain presents density wave order.
We note that a non-reciprocal density-wave pattern has been observed in purely dissipative non-reciprocal bosonic systems~\cite{Rabl2024}.


In the density wave phase correlations decay symmetrically, $\zeta_R = \zeta_L$.
Depending on the nature of the pairing interaction, coherent or non-reciprocal, two opposite behaviors emerge.
Coherent pairing promotes long-range correlations with $\zeta\propto\Delta$, while for $\Gamma_p = 2\Delta$ correlations are extremely short-ranged, yielding saturating $\zeta$ for increasing $\Delta$.

\textit{Conclusion ---}
In this work, we investigated the breakdown of non-reciprocity due to superconducting pairing.
We reveal a phase transition corresponding to a qualitative change in the spectrum of the dynamical non-Hermitian Kitaev matrix.
The non-reciprocal nature of the dominant processes determines the behavior of the system: directional dynamics and inhomogeneous particle density on one hand and pairing-induced density wave order on the other.
The non-reciprocal phase is further characterized by long relaxation times, typical of non-reciprocal systems~\cite{Ueda2019c,Clerk2023} with emergent large lengthscales.
The density wave phase, instead, presents fascinating spatial correlations, whose spreading is tunable by pairing non-reciprocity.

Our results contribute to the understanding of non-reciprocal quantum matter and directly connect with recent studies on non-reciprocal fermionic systems~\cite{Clerk2022,Clerk2023,Brunelli2025}.
In particular, the non-reciprocal Kitaev chain provides a tractable framework to investigate the behavior of non-reciprocal fermions beyond the phase diagram presented in this work, and the non-trivial behavior of correlations invites a more detailed analysis of the effect of non-reciprocal pairing.
Our results are based on the Lindblad master equation, and therefore do not suffer from any postselection limitation, making our model particularly appealing for experimental investigation. 
In this regard, an implementation of our model in quantum dots coupled to common leads~\cite{Nunnenkamp2018} would be of great interest in the ongoing discussion on the so-called poor man's Majoranas~\cite{Flensberg2012,Goswami2024,Dvir2024,Dvir2024a,Goswami2025}.

\begin{acknowledgments}
    \textit{Acknowledgments ---}
    P.~B.~acknowledges support by the Austrian Science Fund (FWF) [Grant Agreement No.~10.55776/ESP9057324].
    This research was funded in whole or in part by the Austrian Science Fund (FWF) [10.55776/COE1].
    For Open Access purposes, the authors have applied a CC BY public copyright license to any author accepted manuscript version arising from this submission.
\end{acknowledgments}

\newpage

\appendix
\begin{center}
    \textbf{End Matter}
\end{center}

\section{Periodic Boundary Conditions}

\begin{figure}[b]
    \centering
    \includegraphics[width=0.99\linewidth]{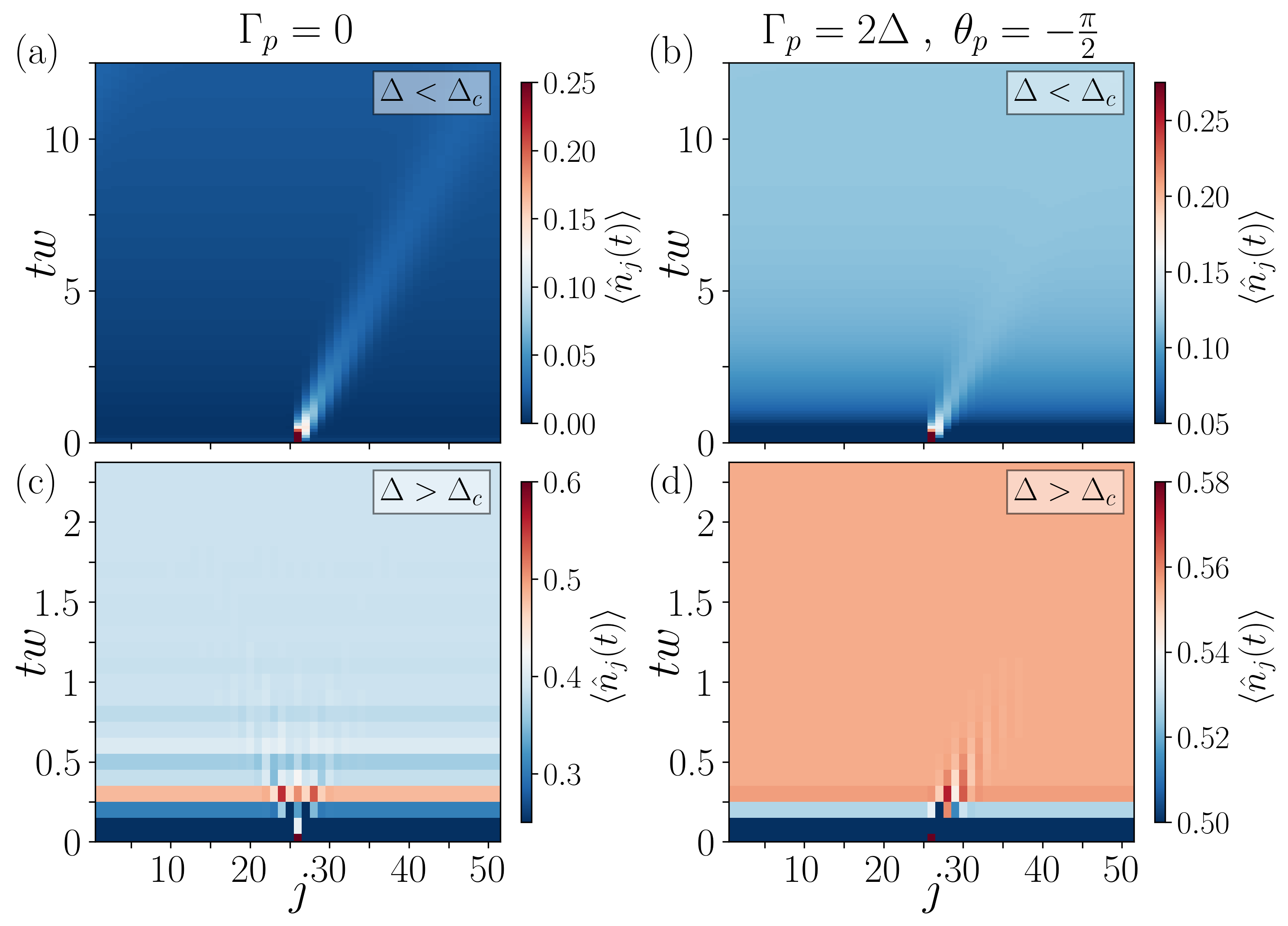}
    \caption{\label{Fig:nt PBC}
    Density dynamics $\langle\hat{n}_j(t)\rangle$ under PBC show a transition similar to the one presented in the main text. 
    (a)-(b) At weak pairing, particles spread only towards the right in a unidirectional lightcone.
    (c)-(d) Above the critical point $\langle\hat{n}_j(t)\rangle$ shows the typical density wave pattern, which becomes non-reciprocal upon making pairing non-reciprocal.
    Notice that the steady state density is completely homogeneous, hence these features concern the transient dynamics.
    }
\end{figure}

In the main part of this work we focused on the behavior of the fermionic Kitaev chain in open boundary conditions.
Here, we report some of the main results under periodic boundary conditions.
The PBC dynamical matrix is a $2\times 2$ matrix
\begin{equation}
    \label{Eq:H dyn k}
    \mathbb{H}_k = \begin{pmatrix}
        -(\xi_k +\imath\Gamma_k^{(h)}) -\imath\Gamma_p & \imath(2\Delta\sin k - \Gamma_k^{(p)})^* \\
        -\imath(2\Delta\sin k + \Gamma_k^{(p)}) & (\xi_k - \imath\Gamma_{-k}^{(h)}) -\imath \Gamma_p
    \end{pmatrix},
\end{equation}
where $\xi_k = -(2w\cos k + \mu)$ is the tight-binding single particle energy and we introduced momentum-dependent dissipation rates $\Gamma^{(h)}_k = \Gamma_h[1+\cos(k+\theta_h)]$, and $\Gamma^{(p)}_k = \Gamma_pe^{\imath\theta_p}\cos k$.
The spectrum of the dynamical matrix $\mathbb{H}_k$ can be obtained analytically and is reported in full generality in the Supplemental Material~\cite{Supplementary}.
The noise term is proportional to a Kronecker delta between different momenta $\mathbb{F}_{pq} \propto\delta_{pq}$, thus yielding a steady state where only same-momentum correlators are involved.

The transition in the dynamics observed in OBC persists under PBC, as shown in Figure~\ref{Fig:nt PBC}.
The different phases present features akin to the ones shown in the main text, including directional dynamics and long relaxation times in the non-reciprocal phase, and density wave pattern and short relaxation times in the density wave phase.
As the steady state has completely homogeneous particle density, these features affect the transient dynamics only.

\begin{figure}[t]
    \centering
    \includegraphics[width=0.99\linewidth]{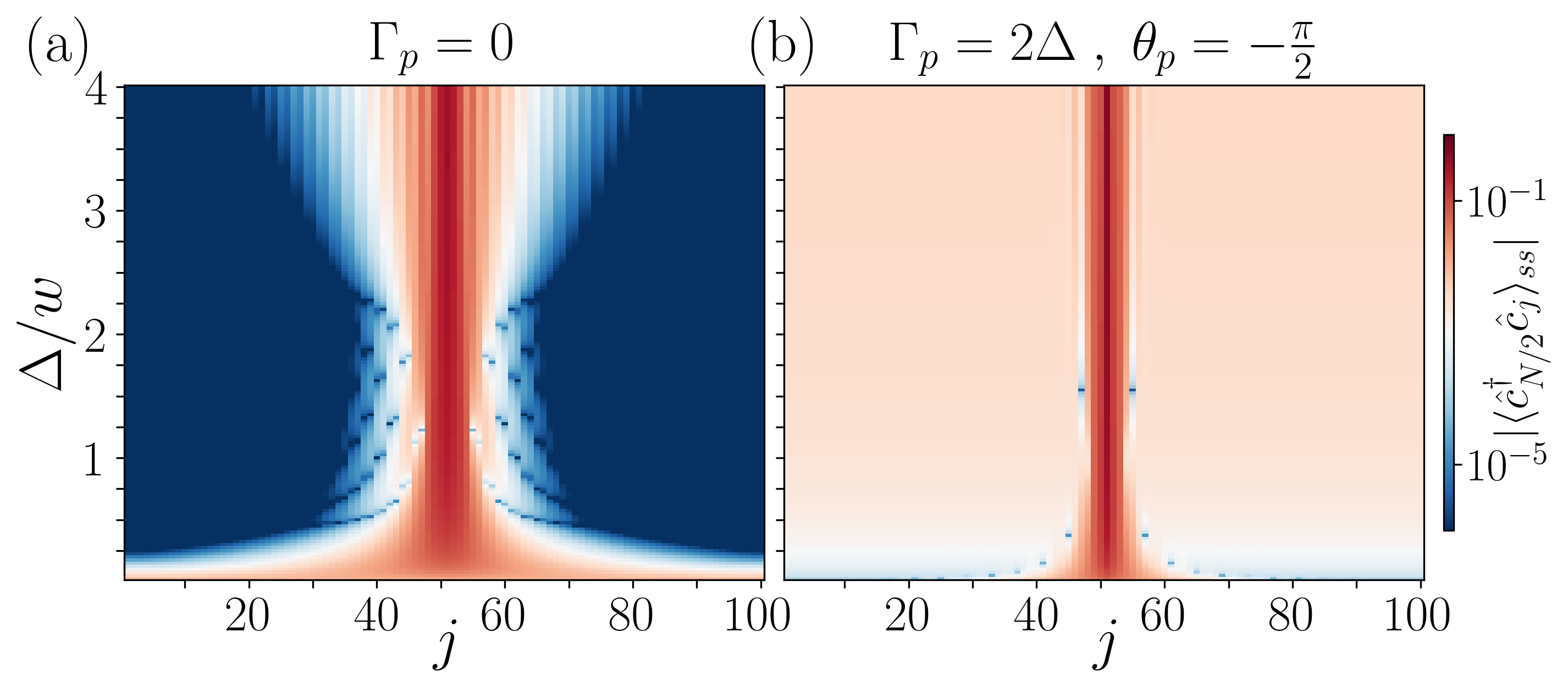}
    \caption{\label{Fig:PBC steady state}
    Single-particle correlation functions $\langle \Cdag{N/2}\C{j}\rangle_{ss}$in PBC across the phase transition.
    (a) In the coherent pairing case, correlations decay slowly away from the critical point, and their range shrinks approaching it both from above and from below.
    (b) For non-reciprocal pairing, correlations have shorter range, and present a finite plateau away from $j=N/2$.
    }
\end{figure}

In the steady state, then, we focus on the correlation functions.
As we show in Figure~\ref{Fig:PBC steady state}, the central site correlations decay similarly to their equivalent in OBC.
Therefore, they carry information about the phase transition in the non-monotonic behavior of the decay length, which decreases for $\Delta \to \Delta_c$ and increases with $\Delta$ in the density wave phase.

\section{System size scaling of relevant lengthscales}

\begin{figure}[t]
    \centering
    \includegraphics[width=0.99\linewidth]{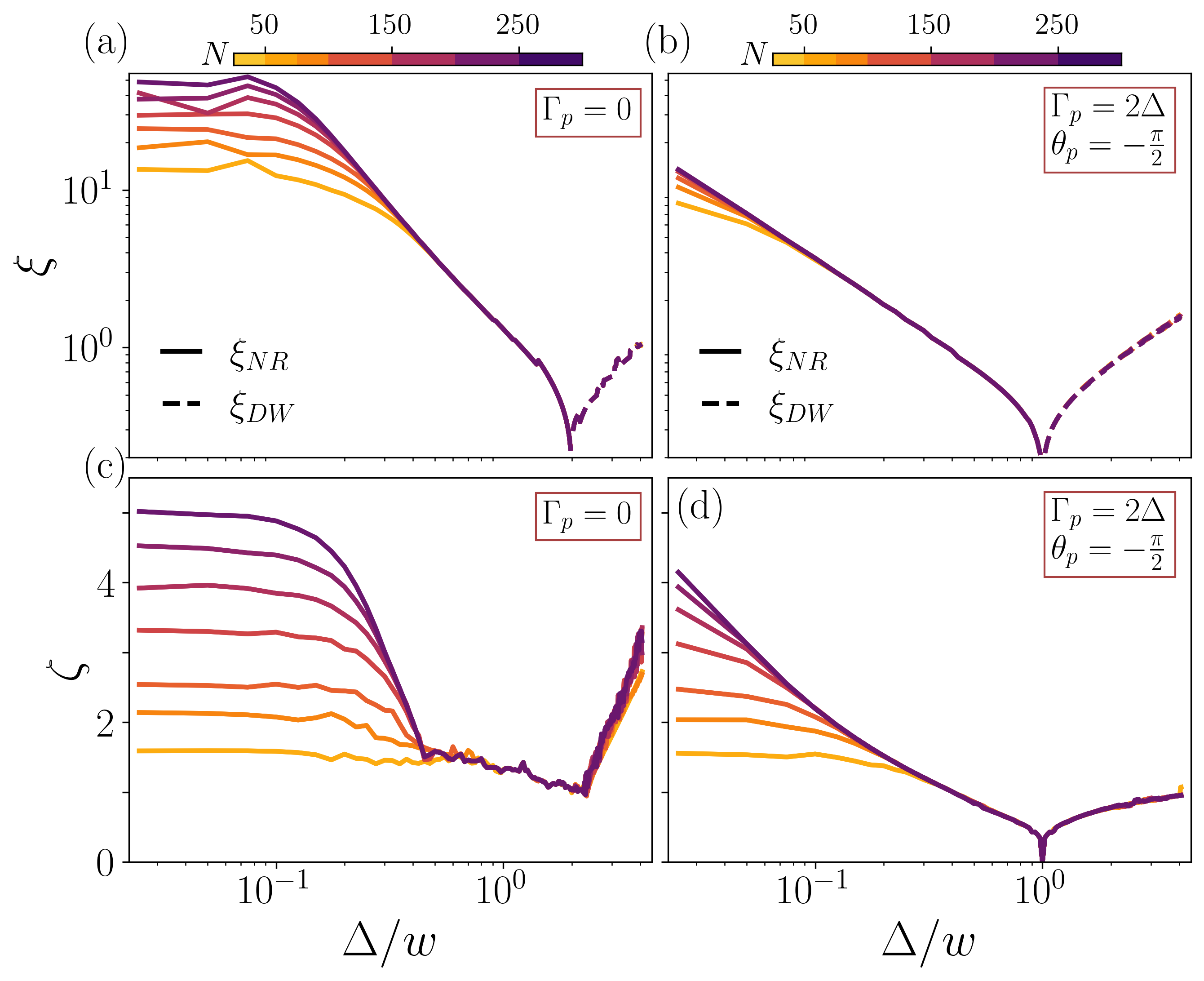}
    \caption{\label{Fig:xi scaling}
    (a)-(b) System size scaling of the density-related lengthscales, $\xi_{NR}$ and $\xi_{DW}$, shown as solid and dashed lines, respectively.
    In the non-reciprocal phase $\xi_{NR}$ diverges as $\Delta\to 0$ and therefore it shows finite size effects whenever $\xi_{NR}\approx N$.
    In the density wave phase, instead, finite size effects are absent.
    (c)-(d) $N$ scaling of the correlations decay length, $\zeta$.
    The non-reciprocal phase is affected by finite $N$, while in the density wave phase $\zeta$ shows very weak dependence on system size.
    }
\end{figure}

In the main text, we introduced several lengthscales describing the various phases in the steady state.
Here, in Figure~\ref{Fig:xi scaling} we briefly discuss their system size scaling.
In panels~(a) and~(b) we show the behavior of the density-related lengthscales $\xi_{NR}$ and $\xi_{DW}$ as solid and dashed lines, respectively.
The non-reciprocal lengthscale, $\xi_{NR}$ presents finite size effects in the weak pairing regime $\Delta\ll w$.
Due to the divergence of $\xi_{NR}$ as $\Delta\to 0$, whenever it approaches the system size $N$, $\xi_{NR}$ deviates from the expected behavior and reaches a plateau with further decreasing $\Delta$.
Due to the faster divergence for $\Gamma_p = 0$ this appears at larger values of $\Delta$ in comparison with the non-reciprocal pairing case.

In panels~(c) and~(d), instead, we show the scaling of the correlations decay length $\zeta$.
Once again, the non-reciprocal phase is affected by strong finite size effects upon vanishing $\Delta$. 
In particular, in the coherent pairing case [Figure~\ref{Fig:xi scaling}(c)] the decay length grows very fast as $\Delta$ decreases below a certain threshold $\Delta/w\approx0.4$, before eventually saturating to an $N$-dependent value.
A similar behavior can be observed in the non-reciprocal pairing case, where however finite size effects are smaller.

In the non-reciprocal phase, the lengthscales defined in the main text are all affected by finite system size, particularly approaching $\Delta = 0$.
Crucially, no significant finite size effect can be observed in the density wave phase.

\clearpage
\newpage
\onecolumngrid
\appendix
\begin{center}
    \textbf{Supplemental Material for ``\mytitle''}

\vspace{0.5cm}

    Pietro Brighi and Andreas Nunnenkamp

\vspace{0.25cm}

    \small{\textit{Faculty of Physics, University of Vienna, Boltzmanngasse 5, 1090 Vienna, Austria}}
\end{center}

\setcounter{page}{1}

\section{A: Details on the equations of motion}
\renewcommand{\theequation}{A\arabic{equation}}
\setcounter{equation}{0}
\renewcommand{\thefigure}{A\arabic{figure}}
\setcounter{figure}{0}

In the main part of the text, we reported the dynamical matrix describing the behavior of correlations in open and periodic boundary conditions (End matter).
Here, we give a more detailed derivation of the equations of motion starting from the Hamiltonian~(\ref{Eq:H}), the jump terms~(\ref{Eq:L}), and the Lindblad master equation for operators
\begin{equation}
    \label{Eq:Lind}
    \frac{d\langle\hat{O}\rangle}{dt} = \imath\langle[\hat{H},\hat{O}]\rangle + \sum_{j,n} \langle\hat{L}_j^{(n)\dagger}\hat{O}\hat{L}^{(n)}_j\rangle - \frac{1}{2}\langle\{\hat{L}^{(n)\dagger}_j\hat{L}^{(n)}_j,\hat{O}\}\rangle.
\end{equation}
Since we deal with a quadratic superconducting fermionic system, we consider $G_{\ell m} = \langle\Cdag{\ell}\C{m}\rangle$ and $F_{\ell m} = \langle\Cdag{\ell}\Cdag{m}\rangle$ ($G_{p q} = \langle\Cdag{p}\C{q}\rangle$ and $F_{pq} = \langle\Cdag{p}\Cdag{1}\rangle$ in PBC) for the equations of motion to close.

\subsection{A1: Open boundary conditions}

\renewcommand{\theequation}{A1.\arabic{equation}}
\setcounter{equation}{0}
\renewcommand{\thefigure}{A1.\arabic{figure}}
\setcounter{figure}{0}

We consider a system with open boundaries consisting of $N$ sites.
In open boundary conditions (OBC) the equation of motion for the Green's function reads 
\begin{equation}
    \label{Eq:EoM Glm}
    \begin{split}
    \dot{G}_{\ell m} &= -\imath ( w_R G_{\ell + 1 m} + w_L G_{\ell - 1 m} - w^*_R G_{\ell m + 1} - w^*_L G_{\ell m - 1} \\
    &+\Delta_R F_{\ell m + 1} - \Delta_L F_{\ell m - 1} + \Delta^*_R F^*_{\ell + 1 m} - \Delta^*_L F^*_{\ell - 1 m})- 2(\Gamma_h + \Gamma_p)G_{\ell m} + \Gamma_p\delta_{\ell m},
    \end{split}
\end{equation}
where we introduced the non-reciprocal hopping terms $w_{R/L} = w - \imath\frac{\Gamma_h}{2}e^{\mp\imath\theta_h}$ and non-reciprocal pairing $\Delta_{R/L} = \Delta \mp \imath\frac{\Gamma_p}{2}e^{\imath\theta_p}$.
We notice that the non-reciprocal hopping and pairing have opposite dependence on the phases $\theta_{h,p}$, therefore, if $\theta_h = \theta_p$ hopping and pairing will be non-reciprocal with opposite directionality. 
As previously mentioned, the equation of motion Eq.~(\ref{Eq:EoM Glm}) alone does not close, and one needs to calculate the time evolution of the anomalous Green's function
\begin{align}
    \notag
    \dot{F}_{\ell m} &= -\imath\bigr[ w_R(F_{\ell + 1 m} + F_{\ell m + 1}) + w_L(F_{\ell - 1 m} + F_{\ell m - 1}) \\
    \notag
    &- \Delta^*_R (G_{\ell m + 1} - G^*_{\ell + 1 m}) + \Delta^*_L(G_{\ell m - 1} - G^*_{\ell - 1 m})\bigr] - 2(\Gamma_h + \Gamma_p + \imath \mu)F_{\ell m} +\imath \Delta^*_R(\delta_{\ell + 1 m} - \delta_{\ell m + 1}).
    \label{Eq:EoM Flm}
\end{align}

Combining the Green's functions into a $2N\times 2N$ correlation matrix 
\begin{equation}
    \mathcal{C} = \left(
    \begin{array}{c|c}
    G_{\ell m} & F_{\ell m} \\
    \hline
    -F^*_{\ell m}  & -G^*_{\ell m}
    \end{array}
    \right)
\end{equation}
we write the equations above in a convenient matrix form, introducing a non-Hermitian dynamical matrix $\mathbb{H}$.
In this formalism, the equations of motion assume a von Neumann-like form
\begin{equation}
    \label{Eq:EoM H}
    \dot{\mathcal{C}} = -\imath \left(\mathbb{H} \mathcal{C} - \mathcal{C}\mathbb{H}^\dagger\right) + \mathbb{F},
\end{equation}
where we introduced the noise term $\mathbb{F}$ generated by coherent and incoherent pairing.
Both the non-Hermitian dynamical matrix and the noise term have distinct expressions in the first and second $N$ rows.
In particular,
\begin{equation}
    \mathbb{H}_{mn} = 
    \begin{cases}
        & w_R\delta_{m n-1} + w_L\delta_{m n+1} - \Delta^*_R\delta_{m n-N-1} + \Delta^*_L\delta_{m n-N +1} - [\imath(\Gamma_h + \Gamma_p) - \mu]\delta_{mn} \;\; m\leq N \\
        & -w^*_R\delta_{m n-1} -w^*_L\delta_{m n+1} + \Delta_R\delta_{m n+N-1} - \Delta_L\delta_{m n+N+1} - [\imath(\Gamma_h + \Gamma_p) + \mu] \delta_{mn} \;\; m > N
    \end{cases},
\end{equation}
which corresponds to a non-Hermitian Kitaev chain where both hopping and pairing can be non-reciprocal.
Therefore, through the engineered dissipation introduced by the jump operators~(\ref{Eq:L}) the non-Hermitian Kitaev chain naturally arises without taking the no-click limit.
The noise term instead reads
\begin{equation}
    \label{Eq:Fixed term}
    \mathbb{F}_{\ell m} = 
    \begin{cases}
        &\Gamma_p\delta_{\ell m} +\imath\Delta^*_R(\delta_{\ell + N + 1 m} - \delta_{\ell +N m + 1}) \;\; \ell\leq N \\
        &-\Gamma_p\delta_{\ell m} +\imath\Delta_R(\delta_{\ell + 1 m + N} - \delta_{\ell m + N + 1}) \;\; \ell > N
    \end{cases}.
\end{equation}
As noticed above, this system has a non-trivial steady state ($\mathbb{F}\neq \mathbb{0}$) if $\Delta \neq 0$ or $\Gamma_p \neq 0$.

\subsection{A2: Periodic boundary conditions}

\renewcommand{\theequation}{A2.\arabic{equation}}
\setcounter{equation}{0}
\renewcommand{\thefigure}{A2.\arabic{figure}}
\setcounter{figure}{0}

When periodic boundary conditions (PBC) are applied, it is convenient to write the equations of motion in terms of the quasi-momentum $k = \frac{2\pi}{N}n$, with $n = 1\dots N$.
Transforming the annihilation and creation operators in momentum space $\C{j} = \frac{1}{\sqrt{N}}\sum_k e^{\imath k j}\C{k}$ the Kitaev Hamiltonian can be written as
\begin{equation}
    \label{Eq:H PBC}
    \hat{H} \!=\! \sum_{k>0}\! \xi_k \! \left(\! \Cdag{k}\C{k} \! + \! \Cdag{-k}\C{-k}\!\right)\! +\! 2\imath\sin k \!\left(\!\Delta\Cdag{k}\Cdag{-k}\! -\! \Delta^* \C{-k}\C{k}\!\right),
\end{equation}
where $\xi_k = -(2w\cos k + \mu)$.
The jump operators become
\begin{align}
    \label{Eq:L1 PBC}
    \hat{L}^{(h)}_j &= \sqrt{\frac{\Gamma_h}{N}}\sum_k e^{\imath k j}\left(1+e^{\imath(\theta_h + k)}\right)\C{k}\\
    \label{Eq:L2 PBC}
    \hat{L}^{(p)}_j &= \sqrt{\frac{\Gamma_p}{N}}\sum_k e^{\imath k j}\C{k} + e^{-\imath k j}e^{\imath(\theta_p - k)}\Cdag{k}.
\end{align}

Using the expressions above, we obtain the equation of motion for the Green's function in momentum space 
\begin{equation}
    \label{Eq:EoM G PBC}
    \begin{split}
   \dot{G}_{pq} \!=\! \left[\imath\left(\xi_p + \imath\Gamma^{(h)}_p\right) \! - \! \imath\left(\xi_q - \imath\Gamma^{(h)}_q\right) \!- \!2\Gamma_p\!\right]\! G_{pq} \!+\! \Gamma_p\delta_{qp}\!+\!\left(2\Delta\sin q - \Gamma^{(p)}_q\right)\!F_{p-q}\! + \!\left(2\Delta\sin p - \Gamma^{(p)}_p\right)^*\!(-F^*_{-pq}).
    \end{split}
\end{equation}
To simplify the notation, we have introduced momentum dependent dissipation rates $\Gamma^{(h)}_k = \Gamma_h(1+\cos(k+\theta_h))$ and $\Gamma^{(p)}_k = \Gamma_pe^{\imath\theta_p}\cos k$.
As in the OBC case, the equation of motion of the Green's function involve the anomalous Green's functions $F_{p-q}$, and $-F^*_{-pq}$ whose equation of motion are obtained by the generic equation of motion for $F_{pq}$
\begin{equation}
    \label{Eq:EoM F PBC}
    \begin{split}
        \dot{F}_{pq} \!=\! \left[\imath\left(\xi_p \!+\! \imath\Gamma^{(h)}_p\right) \!+\! \imath\left(\xi_q \!+\! \Gamma^{(h)}_q\right) \!-2\Gamma_p \!\right]\!F_{pq} \!-\!\left(2\Delta^*\sin p \!-\!\Gamma^{(p)^*}_p\right)\!G_{-pq}^*\!\! +\! \left(2\Delta^*\sin q -\Gamma^{(p)^*}_q\right)\!G_{p-q}\!+\!\left(2\Delta^* \!+\! \imath e^{\imath\theta_p}\Gamma_p\!\right)\!\sin p \delta_{-pq}.
    \end{split}
\end{equation}

The behavior of the Green's function is then fully determined by $4$ coupled differential equations defining $\dot{G}_{pq}$, $\dot{F}_{p-q}$, $-\dot{F}^*_{-pq}$ and $-\dot{G}^*_{-p-q}$.
These equations can be cast in a matrix-matrix form, introducing the momentum space correlation matrix
\begin{equation}
    \label{Eq:C pq}
    \mathcal{C}_{pq} = \begin{pmatrix}
        G_{pq} & F_{p-q} \\
        -F^*_{-pq} & -G^*_{-p-q}
    \end{pmatrix}
\end{equation}
and the coupled equations of motion can be conveniently written as
\begin{equation}
    \label{Eq:EoM PBC H}
    \dot{\mathcal{C}}_{pq} = -\imath(\mathbb{H}_{p}\mathcal{C}_{pq} - \mathcal{C}_{pq}\mathbb{H}^\dagger_{q}) + \mathbb{F}_{pq}
\end{equation}
defining the $2\times 2$ non-Hermitian Bloch matrix $\mathbb{H}_{k}$ and the noise term $\mathbb{F}_{pq}$.
The non-Hermitian Bloch matrix reads
\begin{equation}
    \label{Eq:H dyn k SM}
    \mathbb{H}_k = \begin{pmatrix}
        -(\xi_k +\imath\Gamma_k^{(h)}) -\imath\Gamma_p & \imath(2\Delta\sin k - \Gamma_k^{(p)})^* \\
        -\imath(2\Delta\sin k + \Gamma_k^{(p)}) & (\xi_k - \imath\Gamma_{-k}^{(h)}) -\imath \Gamma_p
    \end{pmatrix}
\end{equation}
and its spectrum can be obtained analytically from the secular equation, giving
\begin{equation}
    \label{Eq:PBC spectrum}
    \begin{split}
    h_{k}^{\pm} &= -\imath\left(\Gamma_h(1+\cos k\cos\theta_h) + \Gamma_p\right) \\
    &\pm\sqrt{\xi_k^2 + 4 \Delta^2\sin^2k - |\Gamma^{(p)}_k|^2 
    + 4\imath\Delta\Gamma_p\sin\theta_p\sin k \cos k - \Gamma_h \sin\theta_h\sin k (2\imath\xi_k + \Gamma_h \sin\theta_h\sin k)}.    
    \end{split}
\end{equation}
The noise term reads
\begin{equation}
    \label{Eq:Fpq}
    \mathbb{F}_{pq} = \begin{pmatrix}
        \Gamma_p & (2\Delta + \imath e^{\imath\theta_p}\Gamma_p)\sin p  \\
        (2\Delta + \imath e^{\imath\theta_p}\Gamma_p)^*\sin p & -\Gamma_p
    \end{pmatrix} \delta_{pq}.
\end{equation}

\section{B: Spectrum of $\mathbb{H}_k$}

\renewcommand{\theequation}{B\arabic{equation}}
\setcounter{equation}{0}
\renewcommand{\thefigure}{B\arabic{figure}}
\setcounter{figure}{0}

\begin{figure}[t]
    \centering
    \includegraphics[width=0.99\linewidth]{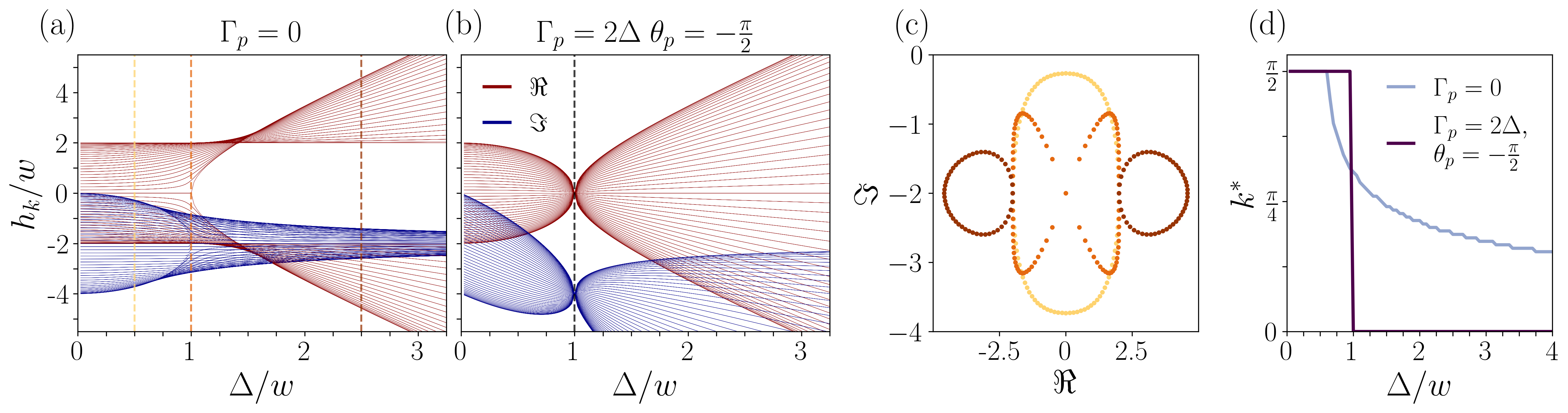}
    \caption{\label{Fig:PBC spectra} 
    Real and imaginary part of the spectrum for PBC as a function of $\Delta/w$ for fixed $ \Gamma_h = 2w$, $\theta_h = \pi/2$ and $\mu = 0$.
    (a) In absence of incoherent pairing, $\Gamma_p = 0$, the spectrum shows a transition at $\Delta = w$ where a gap opens in the real part.
    (b) For non-reciprocal pairing, instead, an $N$-fold exceptional point appears at $\Delta = w$, separating two gapless phases.
    (c) The spectral curve in complex plane for $\Gamma_p = 0$: in the non-reciprocal phase $\Delta = w/2$ (yellow), at the critical point $\Delta = w$ (orange) and in the density wave phase $\Delta = 2.5w$ (red).
    (d) The momentum corresponding to the largest imaginary eigenvalue, $k^*$, as a function of $\Delta/w$ for coherent and non-reciprocal pairing shows a transition from the non-reciprocal phase ($k^* = \pi/2$) to the density wave phase ($k^*<\pi/2$).
    }
\end{figure}

The spectrum Eq.~(\ref{Eq:PBC spectrum}) greatly simplifies for the parameters chosen in the main text, namely $\Gamma_h = 2w,\;\theta_h=\pi/2,\;\Gamma_p = 0,2\Delta$ and $\theta_p = 0,-\pi/2$.
For purely coherent pairing, $\Gamma_p =0$, the PBC spectrum can be written as 
\begin{equation}
    \label{Eq:hk Gp = 0}
    h_k^{\pm} = -2\left[\imath w \pm\sqrt{w^2(\cos k + \imath\sin k)^2 + \Delta^2\sin^2 k})\right]
\end{equation}
and has gapless imaginary part, asymptotically collapsing to $-2w$ as $\Delta\to\infty$, and vanishing for $k=\pi/2$ in the opposite limit $\Delta\to0$.
The real part, on the other hand, presents a gap opening at $\Delta = w$, as shown in Figure~\ref{Fig:PBC spectra}(a).
The eigenvalues in the complex plane, shown in Figure~\ref{Fig:PBC spectra}(c), present different topological features.
In the non-reciprocal phase (yellow curve) the spectrum forms a continuous curve in the complex plane, while two separated \textit{bands} appear in the density wave phase (red curve), separated by a bow tie shaped curve at the critical point (orange curve).
Introducing pairing non-reciprocity, the spectrum in PBC becomes a circle in the complex plane: 
\begin{equation}
\label{Eq:hk Gp=2D}
    h_k^\pm = -2(\imath(w+\Delta) \pm (\cos k + \imath \sin k )\sqrt{(w+\Delta)(w-\Delta)}).
\end{equation}
Similarly to the OBC case discussed in the main text, in this case an exceptional point arises at $\Delta = w$, where the whole spectrum collapses to $h_k^\pm = -2\imath(w+\Delta)$, and separating two gapless phases [Figure~\ref{Fig:PBC spectra}(b)].

From the spectrum of $\mathbb{H}_k$ one can obtain the momentum $k^*$ with largest imaginary part, corresponding to the slowest decaying mode.
In the non-reciprocal pairing case, real and imaginary part of the spectrum swap their momentum dependence at the exceptional point.
In particular, for $\Delta<w$ the imaginary part is $-2(w+\Delta) \pm 2\sin k \sqrt{(w-\Delta)(w+\Delta)}$ and thus has its maximum at $k^* = \pi/2$, highlighting the non-reciprocal nature of dynamics.
In the density wave phase, instead, the maximum is at $k^* = 0$.
This behavior is reported in Figure~\ref{Fig:PBC spectra}(d), together with the behavior of $k^*$ for $\Gamma_p = 0$.
In this latter case, the transition is less sharp: $k^*$ deviates from $\pi/2$ at $\Delta<w$ and does not drop immediately to $0$.

\section{C: PBC steady state}

\renewcommand{\theequation}{C\arabic{equation}}
\setcounter{equation}{0}
\renewcommand{\thefigure}{C\arabic{figure}}
\setcounter{figure}{0}

Since the noise term is non-zero only if $p=q$, only the same-momentum correlation matrix is non-trivial in the steady state.
To obtain the steady state in PBC, then, we fix $p = q \equiv k$ and we set to zero the LHS of Eq.~(\ref{Eq:EoM PBC H}), obtaining a set of equations
\begin{align}
    \label{Eq:nk ss}
    \langle\hat{n}_k\rangle_{ss} &= \frac{\Gamma_p + 4\Delta\sin k\Re[\langle\Cdag{k}\Cdag{-k}\rangle] - 2\Re[\Gamma_k^{(p)}\langle\Cdag{k}\Cdag{-k}\rangle]}{2\left(\Gamma_k^{(h)} + \Gamma_p\right)} \\ 
    \label{Eq:cdagk ss}
    \langle\Cdag{k}\Cdag{-k}\rangle_{ss} &= \frac{2\sin k \left[\Delta_L - \Delta(\langle\hat{n}_k\rangle + \langle\hat{n}_{-k}\rangle) \right] - \Gamma_k^{(p)^*} (\langle\hat{n}_k\rangle - \langle\hat{n}_{-k}\rangle)}{2\left(\Gamma_h(1+\cos k \cos \theta_h) + \Gamma_p - \imath\xi_k\right)}. 
\end{align}

\begin{figure}[t]
    \centering
    \includegraphics[width=0.99\linewidth]{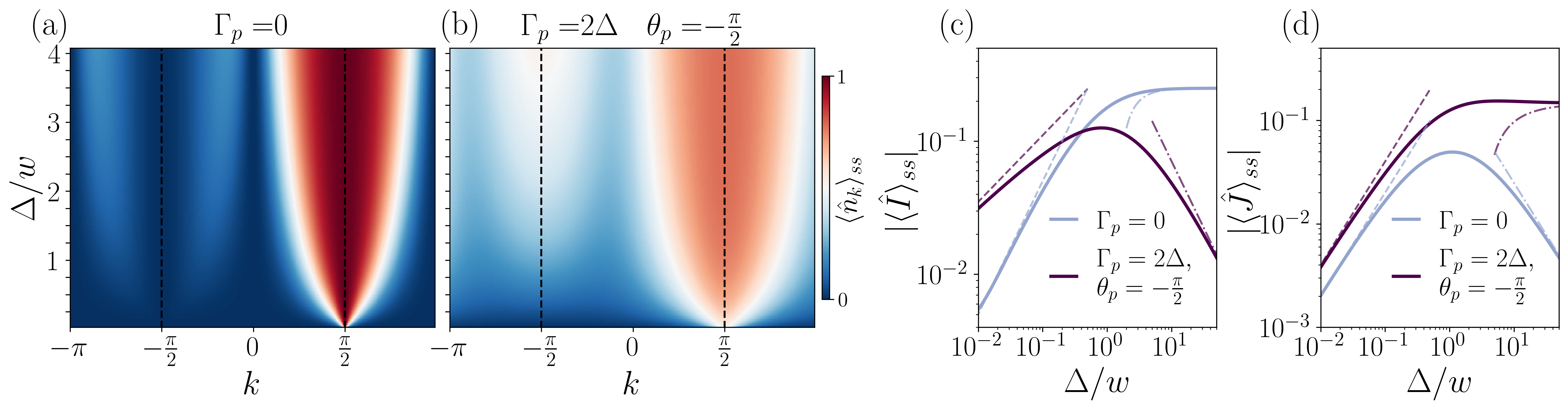}
    \caption{\label{Fig:PBC nk IJ}
    (a)-(b) Momentum occupation across the phase transition for coherent and non-reciprocal pairing.
    In both cases, occupation is initially concentrated around $k=\pi/2$, indicating non-reciprocity.
    However, as $\Delta/w$ increases and one enters the density wave phase, all positive momenta become occupied.
    (c) The current $\hat{I}$ in the steady state shows a clearly different behavior in the two phases. 
    Both for $\Gamma_p = 0$ and $\Gamma_p = 2\Delta$ it increases with $\Delta$ in the non-reciprocal phase, while in the density wave phase it saturates for coherent pairing and it vanishes for non-reciprocal pairing.
    (d) The pairing current $\hat{J}$ also shows signatures of the phase transition, increasing in the non-reciprocal phase, attaining a finite value in the density wave phase for $\Gamma_p = 2\Delta$ and vanishing for $\Gamma_p = 0$.
    }
\end{figure}

Since the steady state correlation matrix in momentum space only has diagonal entries $p=q$, the real space density is always homogeneous irrespective of the pairing amplitude $\Delta$.
Nonetheless, non-reciprocity manifests as a finite current supported by the steady state~\cite{Kawabata2023,Brunelli2025}.
To define the current in our system, we use the real space equations of motion to derive the continuity equation, which is affected by pairing and non-reciprocal hopping.
In particular, for $\Gamma_h = 2w$, $\theta_h = \pi/2$ and $\theta_p = -\pi/2$, the continuity equation reads
\begin{equation}
    \label{Eq:continutiy}
    \partial_t \langle\hat{n}_\ell\rangle =  - 4w\langle\hat{I}_{\ell - 1}\rangle - (2\Delta - \Gamma_p)\langle\hat{J}_{\ell}\rangle +(2\Delta + \Gamma_p)\langle\hat{J}_{\ell - 1}\rangle -2(2w + \Gamma_p)\langle\hat{n}_\ell\rangle + \Gamma_p,
\end{equation}
where $\hat{I}_\ell = \frac{\imath}{2}(\Cdag{\ell}\C{\ell+1} - \Cdag{\ell+1}\C{\ell})$ is the usual current operator and $\hat{J}_\ell = \frac{\imath}{2}(\Cdag{\ell}\Cdag{\ell + 1} - \C{\ell + 1}\C{\ell})$ is the pairing contribution.
The global currents can then be easily evaluated and give $\langle\hat{I}\rangle = \frac{1}{N}\sum_\ell\langle\hat{I}_\ell\rangle = -\frac{1}{N}\sum_k\langle\hat{n}_k\rangle \sin k$, $\langle\hat{J}\rangle = \frac{1}{N}\sum_\ell\langle\hat{J}_\ell\rangle = -\frac{1}{N}\sum_{k>0}\Re[\langle\Cdag{k}\Cdag{-k}\rangle]\sin k$.

Figure~\ref{Fig:PBC nk IJ} shows the behavior of the steady state as a function of $\Delta/w$ across the phase transition identified in the main text.
The phase transition does not affect drastically the steady state momentum occupation, which however smoothly spreads from $k = \pi/2$ to most positive momenta as $\Delta$ increases, indicating the breakdown of non-reciprocity.
For coherent pairing $\Gamma_p = 0$ this results in a linear growth of the current $\langle\hat{I}\rangle_{ss}\propto \Delta/w$ and of the pairing current $\langle\hat{J}\rangle_{ss}\propto \Delta/w$, as shown by the dashed line in panel~(c) and panel~(d), respectively.
As $\Delta$ increases and the system enters the density wave phase $\Delta/w\gg1$, the momentum density spreads over all positive $k$, resulting in a finite current $I_\infty = -\frac{1}{4}$ quickly approached as $|\langle\hat{I}\rangle_{ss}-I_\infty|\propto \left(\frac{\Delta}{t}\right)^{-2}$ [dotted line in panel (c)].
On the other hand, $\langle\hat{J}\rangle_{ss}$ vanishes at large $\Delta/w$, following a $\left(\frac{\Delta}{w}\right)^{-1}$ decay [dotted line in panel (d)].


The situation is somehow reversed in the non-reciprocal pairing scenario $\Gamma_p = 2\Delta$, $\theta_p = -\pi/2$.
The density is still initially peaked around $k = \pi/2$, but with a wider distribution and lower magnitude, resulting in the slower initial growth of the current $\langle\hat{I}\rangle_{ss}\propto\sqrt{\Delta/w}$ shown in panel (c) (dark purple dashed line), while the pairing current increases linearly.
In the density wave phase the momentum density becomes more and more symmetric between positive and negative momenta, leading to a current vanishing as $\left(\Delta/w\right)^{-1}$.
The non-reciprocal nature of pairing, however, induces a finite pairing current in this phase, and $J\to J_\infty = \frac{1}{\sqrt{2}} - 1$ at large $\Delta/w$.

\section{D: System size scaling of critical $\Delta$ from the OBC spectrum}

\renewcommand{\theequation}{D\arabic{equation}}
\setcounter{equation}{0}
\renewcommand{\thefigure}{D\arabic{figure}}
\setcounter{figure}{0}

In the main text, we briefly mentioned the finite size scaling of the critical pairing appearing in the OBC spectrum for $\Gamma_p = 0$.
Here, we provide a detailed analysis of its behavior with system size $N$.

\begin{figure}[h]
    \centering
    \includegraphics[width=0.75\linewidth]{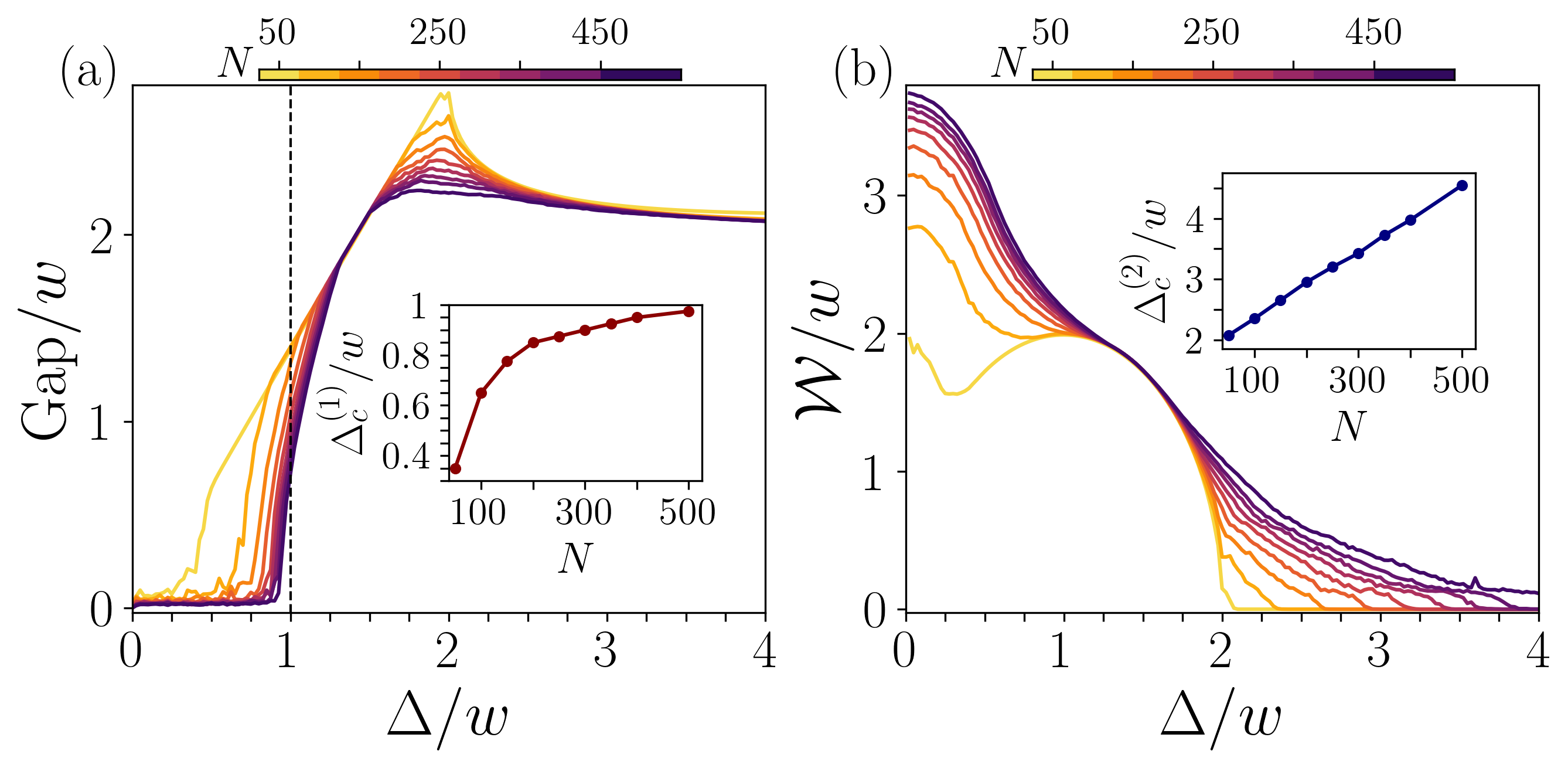}
    \caption{\label{Fig:Dc scaling}
    (a) Gap of the real part of the spectrum in OBC at $\Gamma_p = 0$ for different system sizes .
    In the non-reciprocal phase the spectrum is gapless, while in the density wave phase it attains a finite value $\text{Gap}\approx2w$.
    The critical pairing when the gap opens, $\Delta^{(1)}_c$ approaches $w$ as $N\to\infty$ (inset).
    (b) Bandwidth $\mathcal{W}$ of the imaginary part of the spectrum in OBC at $\Gamma_p = 0$  for different system sizes.
    $\mathcal{W}$ decreases as pairing grows, until eventually vanishes at $\Delta^{(2)}_c$.
    The critical pairing, shown in the inset, grows linearly with system size.
    }
\end{figure}

First, let us analyze the critical value corresponding to the gap opening in the real part of the spectrum.
As observed in Figure~\ref{Fig:spectra}, at $N=100$ the gap opens at $\Delta^{(1)}_c<w$.
In Figure~\ref{Fig:Dc scaling}(a) we show the gap as a function of $\Delta/w$ for different system sizes.
In the non-reciprocal phase, the spectrum is gapless, and at the critical pairing the gap sharply opens, attaining a finite, system size independent, value at $\Delta/w \gg 1$.
In the inset, we show the behavior of $\Delta^{(1)}_c$ as a function of $N$, showing its saturation to $\Delta_c = w$ as $N\to\infty$.

At larger $\Delta$, we noticed in the main text a second critical pairing $\Delta^{(2)}_c$ where the imaginary part of the spectrum becomes completely degenerate.
To further investigate this behavior, we define the bandwidth $\mathcal{W} = \max_n \Im[E_n] - \min_n \Im[E_n]$, whose scaling with $\Delta$ and $N$ is shown in Figure~\ref{Fig:Dc scaling}(b).
Increasing pairing, the bandwidth monotonously decreases (except for small system sizes) and eventually vanishes at $\Delta^{(2)}_c$, indicating a single-valued imaginary part of the spectrum.
In the inset, we show the system size scaling of $\Delta^{(2)}_c$ and observe a clear linear growth with $N$, thus suggesting that the vanishing of the bandwidth is a finite size effect, requiring larger and larger $\Delta$ as $N$ is increased.

\begin{thebibliography}{65}%
\makeatletter
\providecommand \@ifxundefined [1]{%
 \@ifx{#1\undefined}
}%
\providecommand \@ifnum [1]{%
 \ifnum #1\expandafter \@firstoftwo
 \else \expandafter \@secondoftwo
 \fi
}%
\providecommand \@ifx [1]{%
 \ifx #1\expandafter \@firstoftwo
 \else \expandafter \@secondoftwo
 \fi
}%
\providecommand \natexlab [1]{#1}%
\providecommand \enquote  [1]{``#1''}%
\providecommand \bibnamefont  [1]{#1}%
\providecommand \bibfnamefont [1]{#1}%
\providecommand \citenamefont [1]{#1}%
\providecommand \href@noop [0]{\@secondoftwo}%
\providecommand \href [0]{\begingroup \@sanitize@url \@href}%
\providecommand \@href[1]{\@@startlink{#1}\@@href}%
\providecommand \@@href[1]{\endgroup#1\@@endlink}%
\providecommand \@sanitize@url [0]{\catcode `\\12\catcode `\$12\catcode `\&12\catcode `\#12\catcode `\^12\catcode `\_12\catcode `\%12\relax}%
\providecommand \@@startlink[1]{}%
\providecommand \@@endlink[0]{}%
\providecommand \url  [0]{\begingroup\@sanitize@url \@url }%
\providecommand \@url [1]{\endgroup\@href {#1}{\urlprefix }}%
\providecommand \urlprefix  [0]{URL }%
\providecommand \Eprint [0]{\href }%
\providecommand \doibase [0]{https://doi.org/}%
\providecommand \selectlanguage [0]{\@gobble}%
\providecommand \bibinfo  [0]{\@secondoftwo}%
\providecommand \bibfield  [0]{\@secondoftwo}%
\providecommand \translation [1]{[#1]}%
\providecommand \BibitemOpen [0]{}%
\providecommand \bibitemStop [0]{}%
\providecommand \bibitemNoStop [0]{.\EOS\space}%
\providecommand \EOS [0]{\spacefactor3000\relax}%
\providecommand \BibitemShut  [1]{\csname bibitem#1\endcsname}%
\let\auto@bib@innerbib\@empty
\bibitem [{\citenamefont {Fruchart}\ \emph {et~al.}(2021)\citenamefont {Fruchart}, \citenamefont {Hanai}, \citenamefont {Littlewood},\ and\ \citenamefont {Vitelli}}]{Vitelli2021}%
  \BibitemOpen
  \bibfield  {author} {\bibinfo {author} {\bibfnamefont {M.}~\bibnamefont {Fruchart}}, \bibinfo {author} {\bibfnamefont {R.}~\bibnamefont {Hanai}}, \bibinfo {author} {\bibfnamefont {P.~B.}\ \bibnamefont {Littlewood}},\ and\ \bibinfo {author} {\bibfnamefont {V.}~\bibnamefont {Vitelli}},\ }\bibfield  {title} {\bibinfo {title} {Non-reciprocal phase transitions},\ }\href {https://doi.org/10.1038/s41586-021-03375-9} {\bibfield  {journal} {\bibinfo  {journal} {Nature}\ }\textbf {\bibinfo {volume} {592}},\ \bibinfo {pages} {363} (\bibinfo {year} {2021})}\BibitemShut {NoStop}%
\bibitem [{\citenamefont {Avni}\ \emph {et~al.}(2025)\citenamefont {Avni}, \citenamefont {Fruchart}, \citenamefont {Martin}, \citenamefont {Seara},\ and\ \citenamefont {Vitelli}}]{Vitelli2025}%
  \BibitemOpen
  \bibfield  {author} {\bibinfo {author} {\bibfnamefont {Y.}~\bibnamefont {Avni}}, \bibinfo {author} {\bibfnamefont {M.}~\bibnamefont {Fruchart}}, \bibinfo {author} {\bibfnamefont {D.}~\bibnamefont {Martin}}, \bibinfo {author} {\bibfnamefont {D.}~\bibnamefont {Seara}},\ and\ \bibinfo {author} {\bibfnamefont {V.}~\bibnamefont {Vitelli}},\ }\bibfield  {title} {\bibinfo {title} {Nonreciprocal {Ising} model},\ }\href {https://doi.org/10.1103/PhysRevLett.134.117103} {\bibfield  {journal} {\bibinfo  {journal} {Phys. Rev. Lett.}\ }\textbf {\bibinfo {volume} {134}},\ \bibinfo {pages} {117103} (\bibinfo {year} {2025})}\BibitemShut {NoStop}%
\bibitem [{\citenamefont {Belyansky}\ \emph {et~al.}(2025)\citenamefont {Belyansky}, \citenamefont {Weis}, \citenamefont {Hanai}, \citenamefont {Littlewood},\ and\ \citenamefont {Clerk}}]{Clerk2025}%
  \BibitemOpen
  \bibfield  {author} {\bibinfo {author} {\bibfnamefont {R.}~\bibnamefont {Belyansky}}, \bibinfo {author} {\bibfnamefont {C.}~\bibnamefont {Weis}}, \bibinfo {author} {\bibfnamefont {R.}~\bibnamefont {Hanai}}, \bibinfo {author} {\bibfnamefont {P.~B.}\ \bibnamefont {Littlewood}},\ and\ \bibinfo {author} {\bibfnamefont {A.~A.}\ \bibnamefont {Clerk}},\ }\bibfield  {title} {\bibinfo {title} {Phase transitions in nonreciprocal driven-dissipative condensates},\ }\href {https://doi.org/10.1103/gphr-d1bc} {\bibfield  {journal} {\bibinfo  {journal} {Phys. Rev. Lett.}\ }\textbf {\bibinfo {volume} {135}},\ \bibinfo {pages} {123401} (\bibinfo {year} {2025})}\BibitemShut {NoStop}%
\bibitem [{\citenamefont {Soares}\ \emph {et~al.}(2025)\citenamefont {Soares}, \citenamefont {Brunelli},\ and\ \citenamefont {Schirò}}]{Brunelli2025}%
  \BibitemOpen
  \bibfield  {author} {\bibinfo {author} {\bibfnamefont {R.~D.}\ \bibnamefont {Soares}}, \bibinfo {author} {\bibfnamefont {M.}~\bibnamefont {Brunelli}},\ and\ \bibinfo {author} {\bibfnamefont {M.}~\bibnamefont {Schirò}},\ }\href {https://arxiv.org/abs/2505.15711} {\bibinfo {title} {Dissipative phase transition of interacting non-reciprocal fermions}} (\bibinfo {year} {2025}),\ \Eprint {https://arxiv.org/abs/2505.15711} {arXiv:2505.15711 [quant-ph]} \BibitemShut {NoStop}%
\bibitem [{\citenamefont {Yang}\ \emph {et~al.}(2022)\citenamefont {Yang}, \citenamefont {Jiang},\ and\ \citenamefont {Bergholtz}}]{Bergholtz2022}%
  \BibitemOpen
  \bibfield  {author} {\bibinfo {author} {\bibfnamefont {F.}~\bibnamefont {Yang}}, \bibinfo {author} {\bibfnamefont {Q.-D.}\ \bibnamefont {Jiang}},\ and\ \bibinfo {author} {\bibfnamefont {E.~J.}\ \bibnamefont {Bergholtz}},\ }\bibfield  {title} {\bibinfo {title} {Liouvillian skin effect in an exactly solvable model},\ }\href {https://doi.org/10.1103/PhysRevResearch.4.023160} {\bibfield  {journal} {\bibinfo  {journal} {Phys. Rev. Res.}\ }\textbf {\bibinfo {volume} {4}},\ \bibinfo {pages} {023160} (\bibinfo {year} {2022})}\BibitemShut {NoStop}%
\bibitem [{\citenamefont {Begg}\ and\ \citenamefont {Hanai}(2024)}]{Hanai2024}%
  \BibitemOpen
  \bibfield  {author} {\bibinfo {author} {\bibfnamefont {S.~E.}\ \bibnamefont {Begg}}\ and\ \bibinfo {author} {\bibfnamefont {R.}~\bibnamefont {Hanai}},\ }\bibfield  {title} {\bibinfo {title} {Quantum criticality in open quantum spin chains with nonreciprocity},\ }\href {https://doi.org/10.1103/PhysRevLett.132.120401} {\bibfield  {journal} {\bibinfo  {journal} {Phys. Rev. Lett.}\ }\textbf {\bibinfo {volume} {132}},\ \bibinfo {pages} {120401} (\bibinfo {year} {2024})}\BibitemShut {NoStop}%
\bibitem [{\citenamefont {Brighi}\ and\ \citenamefont {Nunnenkamp}(2024)}]{Brighi2024}%
  \BibitemOpen
  \bibfield  {author} {\bibinfo {author} {\bibfnamefont {P.}~\bibnamefont {Brighi}}\ and\ \bibinfo {author} {\bibfnamefont {A.}~\bibnamefont {Nunnenkamp}},\ }\bibfield  {title} {\bibinfo {title} {Nonreciprocal dynamics and the non-{Hermitian} skin effect of repulsively bound pairs},\ }\href {https://doi.org/10.1103/PhysRevA.110.L020201} {\bibfield  {journal} {\bibinfo  {journal} {Phys. Rev. A}\ }\textbf {\bibinfo {volume} {110}},\ \bibinfo {pages} {L020201} (\bibinfo {year} {2024})}\BibitemShut {NoStop}%
\bibitem [{\citenamefont {Porras}\ and\ \citenamefont {Fern\'andez-Lorenzo}(2019)}]{Porras2019}%
  \BibitemOpen
  \bibfield  {author} {\bibinfo {author} {\bibfnamefont {D.}~\bibnamefont {Porras}}\ and\ \bibinfo {author} {\bibfnamefont {S.}~\bibnamefont {Fern\'andez-Lorenzo}},\ }\bibfield  {title} {\bibinfo {title} {Topological amplification in photonic lattices},\ }\href {https://doi.org/10.1103/PhysRevLett.122.143901} {\bibfield  {journal} {\bibinfo  {journal} {Phys. Rev. Lett.}\ }\textbf {\bibinfo {volume} {122}},\ \bibinfo {pages} {143901} (\bibinfo {year} {2019})}\BibitemShut {NoStop}%
\bibitem [{\citenamefont {Wanjura}\ \emph {et~al.}(2020)\citenamefont {Wanjura}, \citenamefont {Brunelli},\ and\ \citenamefont {Nunnenkamp}}]{Wanjura2020}%
  \BibitemOpen
  \bibfield  {author} {\bibinfo {author} {\bibfnamefont {C.~C.}\ \bibnamefont {Wanjura}}, \bibinfo {author} {\bibfnamefont {M.}~\bibnamefont {Brunelli}},\ and\ \bibinfo {author} {\bibfnamefont {A.}~\bibnamefont {Nunnenkamp}},\ }\bibfield  {title} {\bibinfo {title} {Topological framework for directional amplification in driven-dissipative cavity arrays},\ }\href {https://doi.org/10.1038/s41467-020-16863-9} {\bibfield  {journal} {\bibinfo  {journal} {Nature Communications}\ }\textbf {\bibinfo {volume} {11}},\ \bibinfo {pages} {3149} (\bibinfo {year} {2020})}\BibitemShut {NoStop}%
\bibitem [{\citenamefont {Poyatos}\ \emph {et~al.}(1996)\citenamefont {Poyatos}, \citenamefont {Cirac},\ and\ \citenamefont {Zoller}}]{Zoller1996}%
  \BibitemOpen
  \bibfield  {author} {\bibinfo {author} {\bibfnamefont {J.~F.}\ \bibnamefont {Poyatos}}, \bibinfo {author} {\bibfnamefont {J.~I.}\ \bibnamefont {Cirac}},\ and\ \bibinfo {author} {\bibfnamefont {P.}~\bibnamefont {Zoller}},\ }\bibfield  {title} {\bibinfo {title} {Quantum reservoir engineering with laser cooled trapped ions},\ }\href {https://doi.org/10.1103/PhysRevLett.77.4728} {\bibfield  {journal} {\bibinfo  {journal} {Phys. Rev. Lett.}\ }\textbf {\bibinfo {volume} {77}},\ \bibinfo {pages} {4728} (\bibinfo {year} {1996})}\BibitemShut {NoStop}%
\bibitem [{\citenamefont {Metelmann}\ and\ \citenamefont {Clerk}(2015)}]{Clerk2015}%
  \BibitemOpen
  \bibfield  {author} {\bibinfo {author} {\bibfnamefont {A.}~\bibnamefont {Metelmann}}\ and\ \bibinfo {author} {\bibfnamefont {A.~A.}\ \bibnamefont {Clerk}},\ }\bibfield  {title} {\bibinfo {title} {Nonreciprocal photon transmission and amplification via reservoir engineering},\ }\href {https://doi.org/10.1103/PhysRevX.5.021025} {\bibfield  {journal} {\bibinfo  {journal} {Phys. Rev. X}\ }\textbf {\bibinfo {volume} {5}},\ \bibinfo {pages} {021025} (\bibinfo {year} {2015})}\BibitemShut {NoStop}%
\bibitem [{\citenamefont {Hatano}\ and\ \citenamefont {Nelson}(1996)}]{Hatano1996}%
  \BibitemOpen
  \bibfield  {author} {\bibinfo {author} {\bibfnamefont {N.}~\bibnamefont {Hatano}}\ and\ \bibinfo {author} {\bibfnamefont {D.~R.}\ \bibnamefont {Nelson}},\ }\bibfield  {title} {\bibinfo {title} {Localization transitions in non-{Hermitian} quantum mechanics},\ }\href {https://doi.org/10.1103/PhysRevLett.77.570} {\bibfield  {journal} {\bibinfo  {journal} {Phys. Rev. Lett.}\ }\textbf {\bibinfo {volume} {77}},\ \bibinfo {pages} {570} (\bibinfo {year} {1996})}\BibitemShut {NoStop}%
\bibitem [{\citenamefont {McDonald}\ \emph {et~al.}(2022)\citenamefont {McDonald}, \citenamefont {Hanai},\ and\ \citenamefont {Clerk}}]{Clerk2022}%
  \BibitemOpen
  \bibfield  {author} {\bibinfo {author} {\bibfnamefont {A.}~\bibnamefont {McDonald}}, \bibinfo {author} {\bibfnamefont {R.}~\bibnamefont {Hanai}},\ and\ \bibinfo {author} {\bibfnamefont {A.~A.}\ \bibnamefont {Clerk}},\ }\bibfield  {title} {\bibinfo {title} {Nonequilibrium stationary states of quantum non-{Hermitian} lattice models},\ }\href {https://doi.org/10.1103/PhysRevB.105.064302} {\bibfield  {journal} {\bibinfo  {journal} {Phys. Rev. B}\ }\textbf {\bibinfo {volume} {105}},\ \bibinfo {pages} {064302} (\bibinfo {year} {2022})}\BibitemShut {NoStop}%
\bibitem [{\citenamefont {Lee}\ \emph {et~al.}(2023)\citenamefont {Lee}, \citenamefont {McDonald},\ and\ \citenamefont {Clerk}}]{Clerk2023}%
  \BibitemOpen
  \bibfield  {author} {\bibinfo {author} {\bibfnamefont {G.}~\bibnamefont {Lee}}, \bibinfo {author} {\bibfnamefont {A.}~\bibnamefont {McDonald}},\ and\ \bibinfo {author} {\bibfnamefont {A.}~\bibnamefont {Clerk}},\ }\bibfield  {title} {\bibinfo {title} {Anomalously large relaxation times in dissipative lattice models beyond the non-{Hermitian} skin effect},\ }\href {https://doi.org/10.1103/PhysRevB.108.064311} {\bibfield  {journal} {\bibinfo  {journal} {Phys. Rev. B}\ }\textbf {\bibinfo {volume} {108}},\ \bibinfo {pages} {064311} (\bibinfo {year} {2023})}\BibitemShut {NoStop}%
\bibitem [{\citenamefont {Lee}(2016)}]{Lee2016}%
  \BibitemOpen
  \bibfield  {author} {\bibinfo {author} {\bibfnamefont {T.~E.}\ \bibnamefont {Lee}},\ }\bibfield  {title} {\bibinfo {title} {Anomalous edge state in a non-{Hermitian} lattice},\ }\href {https://doi.org/10.1103/PhysRevLett.116.133903} {\bibfield  {journal} {\bibinfo  {journal} {Phys. Rev. Lett.}\ }\textbf {\bibinfo {volume} {116}},\ \bibinfo {pages} {133903} (\bibinfo {year} {2016})}\BibitemShut {NoStop}%
\bibitem [{\citenamefont {Yao}\ and\ \citenamefont {Wang}(2018)}]{Wang2018}%
  \BibitemOpen
  \bibfield  {author} {\bibinfo {author} {\bibfnamefont {S.}~\bibnamefont {Yao}}\ and\ \bibinfo {author} {\bibfnamefont {Z.}~\bibnamefont {Wang}},\ }\bibfield  {title} {\bibinfo {title} {Edge states and topological invariants of non-{Hermitian} systems},\ }\href {https://doi.org/10.1103/PhysRevLett.121.086803} {\bibfield  {journal} {\bibinfo  {journal} {Phys. Rev. Lett.}\ }\textbf {\bibinfo {volume} {121}},\ \bibinfo {pages} {086803} (\bibinfo {year} {2018})}\BibitemShut {NoStop}%
\bibitem [{\citenamefont {Xiujuan~Zhang}\ and\ \citenamefont {Chen}(2022)}]{Chen2022}%
  \BibitemOpen
  \bibfield  {author} {\bibinfo {author} {\bibfnamefont {M.-H.~L.}\ \bibnamefont {Xiujuan~Zhang}, \bibfnamefont {Tian~Zhang}}\ and\ \bibinfo {author} {\bibfnamefont {Y.-F.}\ \bibnamefont {Chen}},\ }\bibfield  {title} {\bibinfo {title} {A review on non-{Hermitian} skin effect},\ }\href {https://doi.org/10.1080/23746149.2022.2109431} {\bibfield  {journal} {\bibinfo  {journal} {Advances in Physics: X}\ }\textbf {\bibinfo {volume} {7}},\ \bibinfo {pages} {2109431} (\bibinfo {year} {2022})}\BibitemShut {NoStop}%
\bibitem [{\citenamefont {Gong}\ \emph {et~al.}(2018)\citenamefont {Gong}, \citenamefont {Ashida}, \citenamefont {Kawabata}, \citenamefont {Takasan}, \citenamefont {Higashikawa},\ and\ \citenamefont {Ueda}}]{Ueda2018}%
  \BibitemOpen
  \bibfield  {author} {\bibinfo {author} {\bibfnamefont {Z.}~\bibnamefont {Gong}}, \bibinfo {author} {\bibfnamefont {Y.}~\bibnamefont {Ashida}}, \bibinfo {author} {\bibfnamefont {K.}~\bibnamefont {Kawabata}}, \bibinfo {author} {\bibfnamefont {K.}~\bibnamefont {Takasan}}, \bibinfo {author} {\bibfnamefont {S.}~\bibnamefont {Higashikawa}},\ and\ \bibinfo {author} {\bibfnamefont {M.}~\bibnamefont {Ueda}},\ }\bibfield  {title} {\bibinfo {title} {Topological phases of non-{Hermitian} systems},\ }\href {https://doi.org/10.1103/PhysRevX.8.031079} {\bibfield  {journal} {\bibinfo  {journal} {Phys. Rev. X}\ }\textbf {\bibinfo {volume} {8}},\ \bibinfo {pages} {031079} (\bibinfo {year} {2018})}\BibitemShut {NoStop}%
\bibitem [{\citenamefont {Kawabata}\ \emph {et~al.}(2019)\citenamefont {Kawabata}, \citenamefont {Shiozaki}, \citenamefont {Ueda},\ and\ \citenamefont {Sato}}]{Ueda2019a}%
  \BibitemOpen
  \bibfield  {author} {\bibinfo {author} {\bibfnamefont {K.}~\bibnamefont {Kawabata}}, \bibinfo {author} {\bibfnamefont {K.}~\bibnamefont {Shiozaki}}, \bibinfo {author} {\bibfnamefont {M.}~\bibnamefont {Ueda}},\ and\ \bibinfo {author} {\bibfnamefont {M.}~\bibnamefont {Sato}},\ }\bibfield  {title} {\bibinfo {title} {Symmetry and topology in non-{Hermitian} physics},\ }\href {https://doi.org/10.1103/PhysRevX.9.041015} {\bibfield  {journal} {\bibinfo  {journal} {Phys. Rev. X}\ }\textbf {\bibinfo {volume} {9}},\ \bibinfo {pages} {041015} (\bibinfo {year} {2019})}\BibitemShut {NoStop}%
\bibitem [{\citenamefont {Kawabata}\ \emph {et~al.}(2022)\citenamefont {Kawabata}, \citenamefont {Shiozaki},\ and\ \citenamefont {Ryu}}]{Ryu2022}%
  \BibitemOpen
  \bibfield  {author} {\bibinfo {author} {\bibfnamefont {K.}~\bibnamefont {Kawabata}}, \bibinfo {author} {\bibfnamefont {K.}~\bibnamefont {Shiozaki}},\ and\ \bibinfo {author} {\bibfnamefont {S.}~\bibnamefont {Ryu}},\ }\bibfield  {title} {\bibinfo {title} {Many-body topology of non-{Hermitian} systems},\ }\href {https://doi.org/10.1103/PhysRevB.105.165137} {\bibfield  {journal} {\bibinfo  {journal} {Phys. Rev. B}\ }\textbf {\bibinfo {volume} {105}},\ \bibinfo {pages} {165137} (\bibinfo {year} {2022})}\BibitemShut {NoStop}%
\bibitem [{\citenamefont {Brunelli}\ \emph {et~al.}(2023)\citenamefont {Brunelli}, \citenamefont {Wanjura},\ and\ \citenamefont {Nunnenkamp}}]{Nunnenkamp2023}%
  \BibitemOpen
  \bibfield  {author} {\bibinfo {author} {\bibfnamefont {M.}~\bibnamefont {Brunelli}}, \bibinfo {author} {\bibfnamefont {C.~C.}\ \bibnamefont {Wanjura}},\ and\ \bibinfo {author} {\bibfnamefont {A.}~\bibnamefont {Nunnenkamp}},\ }\bibfield  {title} {\bibinfo {title} {{Restoration of the non-{Hermitian} bulk-boundary correspondence via topological amplification}},\ }\href {https://doi.org/10.21468/SciPostPhys.15.4.173} {\bibfield  {journal} {\bibinfo  {journal} {SciPost Phys.}\ }\textbf {\bibinfo {volume} {15}},\ \bibinfo {pages} {173} (\bibinfo {year} {2023})}\BibitemShut {NoStop}%
\bibitem [{\citenamefont {Bergholtz}\ \emph {et~al.}(2021)\citenamefont {Bergholtz}, \citenamefont {Budich},\ and\ \citenamefont {Kunst}}]{Bergholtz2021}%
  \BibitemOpen
  \bibfield  {author} {\bibinfo {author} {\bibfnamefont {E.~J.}\ \bibnamefont {Bergholtz}}, \bibinfo {author} {\bibfnamefont {J.~C.}\ \bibnamefont {Budich}},\ and\ \bibinfo {author} {\bibfnamefont {F.~K.}\ \bibnamefont {Kunst}},\ }\bibfield  {title} {\bibinfo {title} {Exceptional topology of non-{Hermitian} systems},\ }\href {https://doi.org/10.1103/RevModPhys.93.015005} {\bibfield  {journal} {\bibinfo  {journal} {Rev. Mod. Phys.}\ }\textbf {\bibinfo {volume} {93}},\ \bibinfo {pages} {015005} (\bibinfo {year} {2021})}\BibitemShut {NoStop}%
\bibitem [{\citenamefont {Zhang}\ \emph {et~al.}(2020)\citenamefont {Zhang}, \citenamefont {Chen}, \citenamefont {Zhang}, \citenamefont {Lang}, \citenamefont {Li},\ and\ \citenamefont {Zhu}}]{Zhu2020}%
  \BibitemOpen
  \bibfield  {author} {\bibinfo {author} {\bibfnamefont {D.-W.}\ \bibnamefont {Zhang}}, \bibinfo {author} {\bibfnamefont {Y.-L.}\ \bibnamefont {Chen}}, \bibinfo {author} {\bibfnamefont {G.-Q.}\ \bibnamefont {Zhang}}, \bibinfo {author} {\bibfnamefont {L.-J.}\ \bibnamefont {Lang}}, \bibinfo {author} {\bibfnamefont {Z.}~\bibnamefont {Li}},\ and\ \bibinfo {author} {\bibfnamefont {S.-L.}\ \bibnamefont {Zhu}},\ }\bibfield  {title} {\bibinfo {title} {Skin superfluid, topological {Mott} insulators, and asymmetric dynamics in an interacting non-{Hermitian} {Aubry-Andr\'e-Harper} model},\ }\href {https://doi.org/10.1103/PhysRevB.101.235150} {\bibfield  {journal} {\bibinfo  {journal} {Phys. Rev. B}\ }\textbf {\bibinfo {volume} {101}},\ \bibinfo {pages} {235150} (\bibinfo {year} {2020})}\BibitemShut {NoStop}%
\bibitem [{\citenamefont {Zhang}\ \emph {et~al.}(2022)\citenamefont {Zhang}, \citenamefont {Denner}, \citenamefont {Bzdušek}, \citenamefont {Sentef},\ and\ \citenamefont {Neupert}}]{Neupert2022}%
  \BibitemOpen
  \bibfield  {author} {\bibinfo {author} {\bibfnamefont {S.-B.}\ \bibnamefont {Zhang}}, \bibinfo {author} {\bibfnamefont {M.~M.}\ \bibnamefont {Denner}}, \bibinfo {author} {\bibfnamefont {T.}~\bibnamefont {Bzdušek}}, \bibinfo {author} {\bibfnamefont {M.~A.}\ \bibnamefont {Sentef}},\ and\ \bibinfo {author} {\bibfnamefont {T.}~\bibnamefont {Neupert}},\ }\bibfield  {title} {\bibinfo {title} {Symmetry breaking and spectral structure of the interacting {Hatano-Nelson} model},\ }\href {https://doi.org/10.1103/PhysRevB.106.L121102} {\bibfield  {journal} {\bibinfo  {journal} {Phys. Rev. B}\ }\textbf {\bibinfo {volume} {106}},\ \bibinfo {pages} {L121102} (\bibinfo {year} {2022})}\BibitemShut {NoStop}%
\bibitem [{\citenamefont {Alsallom}\ \emph {et~al.}(2022)\citenamefont {Alsallom}, \citenamefont {Herviou}, \citenamefont {Yazyev},\ and\ \citenamefont {Brzezi\ifmmode~\acute{n}\else \'{n}\fi{}ska}}]{Brzezinska2022}%
  \BibitemOpen
  \bibfield  {author} {\bibinfo {author} {\bibfnamefont {F.}~\bibnamefont {Alsallom}}, \bibinfo {author} {\bibfnamefont {L.}~\bibnamefont {Herviou}}, \bibinfo {author} {\bibfnamefont {O.~V.}\ \bibnamefont {Yazyev}},\ and\ \bibinfo {author} {\bibfnamefont {M.}~\bibnamefont {Brzezi\ifmmode~\acute{n}\else \'{n}\fi{}ska}},\ }\bibfield  {title} {\bibinfo {title} {Fate of the non-{Hermitian} skin effect in many-body fermionic systems},\ }\href {https://doi.org/10.1103/PhysRevResearch.4.033122} {\bibfield  {journal} {\bibinfo  {journal} {Phys. Rev. Res.}\ }\textbf {\bibinfo {volume} {4}},\ \bibinfo {pages} {033122} (\bibinfo {year} {2022})}\BibitemShut {NoStop}%
\bibitem [{\citenamefont {Mao}\ \emph {et~al.}(2023)\citenamefont {Mao}, \citenamefont {Hao},\ and\ \citenamefont {Pan}}]{Pan2023}%
  \BibitemOpen
  \bibfield  {author} {\bibinfo {author} {\bibfnamefont {L.}~\bibnamefont {Mao}}, \bibinfo {author} {\bibfnamefont {Y.}~\bibnamefont {Hao}},\ and\ \bibinfo {author} {\bibfnamefont {L.}~\bibnamefont {Pan}},\ }\bibfield  {title} {\bibinfo {title} {Non-{Hermitian} skin effect in a one-dimensional interacting {Bose} gas},\ }\href {https://doi.org/10.1103/PhysRevA.107.043315} {\bibfield  {journal} {\bibinfo  {journal} {Phys. Rev. A}\ }\textbf {\bibinfo {volume} {107}},\ \bibinfo {pages} {043315} (\bibinfo {year} {2023})}\BibitemShut {NoStop}%
\bibitem [{\citenamefont {Longhi}(2023)}]{Longhi2023}%
  \BibitemOpen
  \bibfield  {author} {\bibinfo {author} {\bibfnamefont {S.}~\bibnamefont {Longhi}},\ }\bibfield  {title} {\bibinfo {title} {Spectral structure and doublon dissociation in the two-particle non-{Hermitian} {Hubbard} model},\ }\href {https://doi.org/https://doi.org/10.1002/andp.202300291} {\bibfield  {journal} {\bibinfo  {journal} {Annalen der Physik}\ }\textbf {\bibinfo {volume} {535}},\ \bibinfo {pages} {2300291} (\bibinfo {year} {2023})}\BibitemShut {NoStop}%
\bibitem [{\citenamefont {Zheng}\ \emph {et~al.}(2024)\citenamefont {Zheng}, \citenamefont {Qiao}, \citenamefont {Wang}, \citenamefont {Cao},\ and\ \citenamefont {Chen}}]{Chen2024}%
  \BibitemOpen
  \bibfield  {author} {\bibinfo {author} {\bibfnamefont {M.}~\bibnamefont {Zheng}}, \bibinfo {author} {\bibfnamefont {Y.}~\bibnamefont {Qiao}}, \bibinfo {author} {\bibfnamefont {Y.}~\bibnamefont {Wang}}, \bibinfo {author} {\bibfnamefont {J.}~\bibnamefont {Cao}},\ and\ \bibinfo {author} {\bibfnamefont {S.}~\bibnamefont {Chen}},\ }\bibfield  {title} {\bibinfo {title} {Exact solution of the {Bose-Hubbard} model with unidirectional hopping},\ }\href {https://doi.org/10.1103/PhysRevLett.132.086502} {\bibfield  {journal} {\bibinfo  {journal} {Phys. Rev. Lett.}\ }\textbf {\bibinfo {volume} {132}},\ \bibinfo {pages} {086502} (\bibinfo {year} {2024})}\BibitemShut {NoStop}%
\bibitem [{\citenamefont {Ekman}\ and\ \citenamefont {Bergholtz}(2024)}]{Bergholtz2024}%
  \BibitemOpen
  \bibfield  {author} {\bibinfo {author} {\bibfnamefont {C.}~\bibnamefont {Ekman}}\ and\ \bibinfo {author} {\bibfnamefont {E.~J.}\ \bibnamefont {Bergholtz}},\ }\bibfield  {title} {\bibinfo {title} {Liouvillian skin effects and fragmented condensates in an integrable dissipative bose-hubbard model},\ }\href {https://doi.org/10.1103/PhysRevResearch.6.L032067} {\bibfield  {journal} {\bibinfo  {journal} {Phys. Rev. Res.}\ }\textbf {\bibinfo {volume} {6}},\ \bibinfo {pages} {L032067} (\bibinfo {year} {2024})}\BibitemShut {NoStop}%
\bibitem [{\citenamefont {Shen}\ \emph {et~al.}(2024)\citenamefont {Shen}, \citenamefont {Chen}, \citenamefont {Yang},\ and\ \citenamefont {Lee}}]{Lee2024}%
  \BibitemOpen
  \bibfield  {author} {\bibinfo {author} {\bibfnamefont {R.}~\bibnamefont {Shen}}, \bibinfo {author} {\bibfnamefont {T.}~\bibnamefont {Chen}}, \bibinfo {author} {\bibfnamefont {B.}~\bibnamefont {Yang}},\ and\ \bibinfo {author} {\bibfnamefont {C.~H.}\ \bibnamefont {Lee}},\ }\bibfield  {title} {\bibinfo {title} {Observation of the non-{{Hermitian}} skin effect and {Fermi} skin on a digital quantum computer},\ }\href {https://doi.org/10.1038/s41467-025-55953-4} {\bibfield  {journal} {\bibinfo  {journal} {Nature Communications}\ }\textbf {\bibinfo {volume} {16}},\ \bibinfo {pages} {1340} (\bibinfo {year} {2024})}\BibitemShut {NoStop}%
\bibitem [{\citenamefont {Qin}\ and\ \citenamefont {Li}(2024)}]{Li2024}%
  \BibitemOpen
  \bibfield  {author} {\bibinfo {author} {\bibfnamefont {Y.}~\bibnamefont {Qin}}\ and\ \bibinfo {author} {\bibfnamefont {L.}~\bibnamefont {Li}},\ }\bibfield  {title} {\bibinfo {title} {Occupation-dependent particle separation in one-dimensional non-{Hermitian} lattices},\ }\href {https://doi.org/10.1103/PhysRevLett.132.096501} {\bibfield  {journal} {\bibinfo  {journal} {Phys. Rev. Lett.}\ }\textbf {\bibinfo {volume} {132}},\ \bibinfo {pages} {096501} (\bibinfo {year} {2024})}\BibitemShut {NoStop}%
\bibitem [{\citenamefont {Kitaev}(2001)}]{Kitaev2003}%
  \BibitemOpen
  \bibfield  {author} {\bibinfo {author} {\bibfnamefont {A.~Y.}\ \bibnamefont {Kitaev}},\ }\bibfield  {title} {\bibinfo {title} {Unpaired {Majorana} fermions in quantum wires},\ }\href {https://doi.org/10.1070/1063-7869/44/10S/S29} {\bibfield  {journal} {\bibinfo  {journal} {Physics-Uspekhi}\ }\textbf {\bibinfo {volume} {44}},\ \bibinfo {pages} {131} (\bibinfo {year} {2001})}\BibitemShut {NoStop}%
\bibitem [{\citenamefont {McDonald}\ \emph {et~al.}(2018)\citenamefont {McDonald}, \citenamefont {Pereg-Barnea},\ and\ \citenamefont {Clerk}}]{Clerk2018}%
  \BibitemOpen
  \bibfield  {author} {\bibinfo {author} {\bibfnamefont {A.}~\bibnamefont {McDonald}}, \bibinfo {author} {\bibfnamefont {T.}~\bibnamefont {Pereg-Barnea}},\ and\ \bibinfo {author} {\bibfnamefont {A.~A.}\ \bibnamefont {Clerk}},\ }\bibfield  {title} {\bibinfo {title} {Phase-dependent chiral transport and effective non-{Hermitian} dynamics in a bosonic {Kitaev}-{Majorana} chain},\ }\href {https://doi.org/10.1103/PhysRevX.8.041031} {\bibfield  {journal} {\bibinfo  {journal} {Phys. Rev. X}\ }\textbf {\bibinfo {volume} {8}},\ \bibinfo {pages} {041031} (\bibinfo {year} {2018})}\BibitemShut {NoStop}%
\bibitem [{\citenamefont {Wanjura}\ \emph {et~al.}(2023)\citenamefont {Wanjura}, \citenamefont {Slim}, \citenamefont {del Pino}, \citenamefont {Brunelli}, \citenamefont {Verhagen},\ and\ \citenamefont {Nunnenkamp}}]{Nunnenkamp2023BKC}%
  \BibitemOpen
  \bibfield  {author} {\bibinfo {author} {\bibfnamefont {C.~C.}\ \bibnamefont {Wanjura}}, \bibinfo {author} {\bibfnamefont {J.~J.}\ \bibnamefont {Slim}}, \bibinfo {author} {\bibfnamefont {J.}~\bibnamefont {del Pino}}, \bibinfo {author} {\bibfnamefont {M.}~\bibnamefont {Brunelli}}, \bibinfo {author} {\bibfnamefont {E.}~\bibnamefont {Verhagen}},\ and\ \bibinfo {author} {\bibfnamefont {A.}~\bibnamefont {Nunnenkamp}},\ }\bibfield  {title} {\bibinfo {title} {{Quadrature nonreciprocity in bosonic networks without breaking time-reversal symmetry}},\ }\href {https://doi.org/10.1038/s41567-023-02128-x} {\bibfield  {journal} {\bibinfo  {journal} {Nature Physics}\ }\textbf {\bibinfo {volume} {19}},\ \bibinfo {pages} {1429} (\bibinfo {year} {2023})}\BibitemShut {NoStop}%
\bibitem [{\citenamefont {Slim}\ \emph {et~al.}(2024)\citenamefont {Slim}, \citenamefont {Wanjura}, \citenamefont {Brunelli}, \citenamefont {del Pino}, \citenamefont {Nunnenkamp},\ and\ \citenamefont {Verhagen}}]{Verhagen2024}%
  \BibitemOpen
  \bibfield  {author} {\bibinfo {author} {\bibfnamefont {J.~J.}\ \bibnamefont {Slim}}, \bibinfo {author} {\bibfnamefont {C.~C.}\ \bibnamefont {Wanjura}}, \bibinfo {author} {\bibfnamefont {M.}~\bibnamefont {Brunelli}}, \bibinfo {author} {\bibfnamefont {J.}~\bibnamefont {del Pino}}, \bibinfo {author} {\bibfnamefont {A.}~\bibnamefont {Nunnenkamp}},\ and\ \bibinfo {author} {\bibfnamefont {E.}~\bibnamefont {Verhagen}},\ }\bibfield  {title} {\bibinfo {title} {Optomechanical realization of the bosonic {Kitaev} chain},\ }\href {https://doi.org/10.1038/s41586-024-07174-w} {\bibfield  {journal} {\bibinfo  {journal} {Nature}\ }\textbf {\bibinfo {volume} {627}},\ \bibinfo {pages} {767} (\bibinfo {year} {2024})}\BibitemShut {NoStop}%
\bibitem [{\citenamefont {Busnaina}\ \emph {et~al.}(2024)\citenamefont {Busnaina}, \citenamefont {Shi}, \citenamefont {McDonald}, \citenamefont {Dubyna}, \citenamefont {Nsanzineza}, \citenamefont {Hung}, \citenamefont {Chang}, \citenamefont {Clerk},\ and\ \citenamefont {Wilson}}]{Wilson2024}%
  \BibitemOpen
  \bibfield  {author} {\bibinfo {author} {\bibfnamefont {J.~H.}\ \bibnamefont {Busnaina}}, \bibinfo {author} {\bibfnamefont {Z.}~\bibnamefont {Shi}}, \bibinfo {author} {\bibfnamefont {A.}~\bibnamefont {McDonald}}, \bibinfo {author} {\bibfnamefont {D.}~\bibnamefont {Dubyna}}, \bibinfo {author} {\bibfnamefont {I.}~\bibnamefont {Nsanzineza}}, \bibinfo {author} {\bibfnamefont {J.~S.~C.}\ \bibnamefont {Hung}}, \bibinfo {author} {\bibfnamefont {C.~W.~S.}\ \bibnamefont {Chang}}, \bibinfo {author} {\bibfnamefont {A.~A.}\ \bibnamefont {Clerk}},\ and\ \bibinfo {author} {\bibfnamefont {C.~M.}\ \bibnamefont {Wilson}},\ }\bibfield  {title} {\bibinfo {title} {Quantum simulation of the bosonic {Kitaev} chain},\ }\href {https://doi.org/10.1038/s41467-024-47186-8} {\bibfield  {journal} {\bibinfo  {journal} {Nature Communications}\ }\textbf {\bibinfo {volume} {15}},\ \bibinfo {pages} {3065} (\bibinfo {year} {2024})}\BibitemShut {NoStop}%
\bibitem [{\citenamefont {Ezawa}(2019)}]{Ezawa2019}%
  \BibitemOpen
  \bibfield  {author} {\bibinfo {author} {\bibfnamefont {M.}~\bibnamefont {Ezawa}},\ }\bibfield  {title} {\bibinfo {title} {Braiding of {Majorana}-like corner states in electric circuits and its non-{Hermitian} generalization},\ }\href {https://doi.org/10.1103/PhysRevB.100.045407} {\bibfield  {journal} {\bibinfo  {journal} {Phys. Rev. B}\ }\textbf {\bibinfo {volume} {100}},\ \bibinfo {pages} {045407} (\bibinfo {year} {2019})}\BibitemShut {NoStop}%
\bibitem [{\citenamefont {Zhao}\ \emph {et~al.}(2021)\citenamefont {Zhao}, \citenamefont {Guo}, \citenamefont {Kou}, \citenamefont {Zhuang},\ and\ \citenamefont {Liu}}]{Liu2021}%
  \BibitemOpen
  \bibfield  {author} {\bibinfo {author} {\bibfnamefont {X.-M.}\ \bibnamefont {Zhao}}, \bibinfo {author} {\bibfnamefont {C.-X.}\ \bibnamefont {Guo}}, \bibinfo {author} {\bibfnamefont {S.-P.}\ \bibnamefont {Kou}}, \bibinfo {author} {\bibfnamefont {L.}~\bibnamefont {Zhuang}},\ and\ \bibinfo {author} {\bibfnamefont {W.-M.}\ \bibnamefont {Liu}},\ }\bibfield  {title} {\bibinfo {title} {Defective {Majorana} zero modes in a non-{Hermitian} {Kitaev} chain},\ }\href {https://doi.org/10.1103/PhysRevB.104.205131} {\bibfield  {journal} {\bibinfo  {journal} {Phys. Rev. B}\ }\textbf {\bibinfo {volume} {104}},\ \bibinfo {pages} {205131} (\bibinfo {year} {2021})}\BibitemShut {NoStop}%
\bibitem [{\citenamefont {Wang}\ \emph {et~al.}(2021)\citenamefont {Wang}, \citenamefont {Xu}, \citenamefont {Li}, \citenamefont {Xu},\ and\ \citenamefont {Wang}}]{Wang2021}%
  \BibitemOpen
  \bibfield  {author} {\bibinfo {author} {\bibfnamefont {Z.-H.}\ \bibnamefont {Wang}}, \bibinfo {author} {\bibfnamefont {F.}~\bibnamefont {Xu}}, \bibinfo {author} {\bibfnamefont {L.}~\bibnamefont {Li}}, \bibinfo {author} {\bibfnamefont {D.-H.}\ \bibnamefont {Xu}},\ and\ \bibinfo {author} {\bibfnamefont {B.}~\bibnamefont {Wang}},\ }\bibfield  {title} {\bibinfo {title} {Unconventional real-complex spectral transition and {Majorana} zero modes in nonreciprocal quasicrystals},\ }\href {https://doi.org/10.1103/PhysRevB.104.174501} {\bibfield  {journal} {\bibinfo  {journal} {Phys. Rev. B}\ }\textbf {\bibinfo {volume} {104}},\ \bibinfo {pages} {174501} (\bibinfo {year} {2021})}\BibitemShut {NoStop}%
\bibitem [{\citenamefont {Arouca}\ \emph {et~al.}(2023)\citenamefont {Arouca}, \citenamefont {Cayao},\ and\ \citenamefont {Black-Schaffer}}]{Black-Schaffer2023}%
  \BibitemOpen
  \bibfield  {author} {\bibinfo {author} {\bibfnamefont {R.}~\bibnamefont {Arouca}}, \bibinfo {author} {\bibfnamefont {J.}~\bibnamefont {Cayao}},\ and\ \bibinfo {author} {\bibfnamefont {A.~M.}\ \bibnamefont {Black-Schaffer}},\ }\bibfield  {title} {\bibinfo {title} {Topological superconductivity enhanced by exceptional points},\ }\href {https://doi.org/10.1103/PhysRevB.108.L060506} {\bibfield  {journal} {\bibinfo  {journal} {Phys. Rev. B}\ }\textbf {\bibinfo {volume} {108}},\ \bibinfo {pages} {L060506} (\bibinfo {year} {2023})}\BibitemShut {NoStop}%
\bibitem [{\citenamefont {Wang}\ \emph {et~al.}(2015)\citenamefont {Wang}, \citenamefont {Liu}, \citenamefont {Xiong},\ and\ \citenamefont {Tong}}]{Tong2015}%
  \BibitemOpen
  \bibfield  {author} {\bibinfo {author} {\bibfnamefont {X.}~\bibnamefont {Wang}}, \bibinfo {author} {\bibfnamefont {T.}~\bibnamefont {Liu}}, \bibinfo {author} {\bibfnamefont {Y.}~\bibnamefont {Xiong}},\ and\ \bibinfo {author} {\bibfnamefont {P.}~\bibnamefont {Tong}},\ }\bibfield  {title} {\bibinfo {title} {Spontaneous $\mathcal{PT}$-symmetry breaking in non-{Hermitian} {Kitaev} and extended {Kitaev} models},\ }\href {https://doi.org/10.1103/PhysRevA.92.012116} {\bibfield  {journal} {\bibinfo  {journal} {Phys. Rev. A}\ }\textbf {\bibinfo {volume} {92}},\ \bibinfo {pages} {012116} (\bibinfo {year} {2015})}\BibitemShut {NoStop}%
\bibitem [{\citenamefont {Zeng}\ \emph {et~al.}(2016)\citenamefont {Zeng}, \citenamefont {Zhu}, \citenamefont {Chen}, \citenamefont {You},\ and\ \citenamefont {L\"u}}]{Lu2016}%
  \BibitemOpen
  \bibfield  {author} {\bibinfo {author} {\bibfnamefont {Q.-B.}\ \bibnamefont {Zeng}}, \bibinfo {author} {\bibfnamefont {B.}~\bibnamefont {Zhu}}, \bibinfo {author} {\bibfnamefont {S.}~\bibnamefont {Chen}}, \bibinfo {author} {\bibfnamefont {L.}~\bibnamefont {You}},\ and\ \bibinfo {author} {\bibfnamefont {R.}~\bibnamefont {L\"u}},\ }\bibfield  {title} {\bibinfo {title} {Non-{Hermitian} {Kitaev} chain with complex on-site potentials},\ }\href {https://doi.org/10.1103/PhysRevA.94.022119} {\bibfield  {journal} {\bibinfo  {journal} {Phys. Rev. A}\ }\textbf {\bibinfo {volume} {94}},\ \bibinfo {pages} {022119} (\bibinfo {year} {2016})}\BibitemShut {NoStop}%
\bibitem [{\citenamefont {Yuce}(2016)}]{Yuce2016}%
  \BibitemOpen
  \bibfield  {author} {\bibinfo {author} {\bibfnamefont {C.}~\bibnamefont {Yuce}},\ }\bibfield  {title} {\bibinfo {title} {{Majorana} edge modes with gain and loss},\ }\href {https://doi.org/10.1103/PhysRevA.93.062130} {\bibfield  {journal} {\bibinfo  {journal} {Phys. Rev. A}\ }\textbf {\bibinfo {volume} {93}},\ \bibinfo {pages} {062130} (\bibinfo {year} {2016})}\BibitemShut {NoStop}%
\bibitem [{\citenamefont {Klett}\ \emph {et~al.}(2017)\citenamefont {Klett}, \citenamefont {Cartarius}, \citenamefont {Dast}, \citenamefont {Main},\ and\ \citenamefont {Wunner}}]{Wunner2017}%
  \BibitemOpen
  \bibfield  {author} {\bibinfo {author} {\bibfnamefont {M.}~\bibnamefont {Klett}}, \bibinfo {author} {\bibfnamefont {H.}~\bibnamefont {Cartarius}}, \bibinfo {author} {\bibfnamefont {D.}~\bibnamefont {Dast}}, \bibinfo {author} {\bibfnamefont {J.}~\bibnamefont {Main}},\ and\ \bibinfo {author} {\bibfnamefont {G.}~\bibnamefont {Wunner}},\ }\bibfield  {title} {\bibinfo {title} {Relation between $\mathcal{PT}$-symmetry breaking and topologically nontrivial phases in the {Su-Schrieffer-Heeger} and {Kitaev} models},\ }\href {https://doi.org/10.1103/PhysRevA.95.053626} {\bibfield  {journal} {\bibinfo  {journal} {Phys. Rev. A}\ }\textbf {\bibinfo {volume} {95}},\ \bibinfo {pages} {053626} (\bibinfo {year} {2017})}\BibitemShut {NoStop}%
\bibitem [{\citenamefont {Kawabata}\ \emph {et~al.}(2018)\citenamefont {Kawabata}, \citenamefont {Ashida}, \citenamefont {Katsura},\ and\ \citenamefont {Ueda}}]{Ueda2018b}%
  \BibitemOpen
  \bibfield  {author} {\bibinfo {author} {\bibfnamefont {K.}~\bibnamefont {Kawabata}}, \bibinfo {author} {\bibfnamefont {Y.}~\bibnamefont {Ashida}}, \bibinfo {author} {\bibfnamefont {H.}~\bibnamefont {Katsura}},\ and\ \bibinfo {author} {\bibfnamefont {M.}~\bibnamefont {Ueda}},\ }\bibfield  {title} {\bibinfo {title} {Parity-time-symmetric topological superconductor},\ }\href {https://doi.org/10.1103/PhysRevB.98.085116} {\bibfield  {journal} {\bibinfo  {journal} {Phys. Rev. B}\ }\textbf {\bibinfo {volume} {98}},\ \bibinfo {pages} {085116} (\bibinfo {year} {2018})}\BibitemShut {NoStop}%
\bibitem [{\citenamefont {Shibata}\ and\ \citenamefont {Katsura}(2019)}]{Katsura2019}%
  \BibitemOpen
  \bibfield  {author} {\bibinfo {author} {\bibfnamefont {N.}~\bibnamefont {Shibata}}\ and\ \bibinfo {author} {\bibfnamefont {H.}~\bibnamefont {Katsura}},\ }\bibfield  {title} {\bibinfo {title} {Dissipative spin chain as a non-{Hermitian} {Kitaev} ladder},\ }\href {https://doi.org/10.1103/PhysRevB.99.174303} {\bibfield  {journal} {\bibinfo  {journal} {Phys. Rev. B}\ }\textbf {\bibinfo {volume} {99}},\ \bibinfo {pages} {174303} (\bibinfo {year} {2019})}\BibitemShut {NoStop}%
\bibitem [{\citenamefont {Sayyad}\ and\ \citenamefont {Lado}(2023)}]{Lado2023}%
  \BibitemOpen
  \bibfield  {author} {\bibinfo {author} {\bibfnamefont {S.}~\bibnamefont {Sayyad}}\ and\ \bibinfo {author} {\bibfnamefont {J.~L.}\ \bibnamefont {Lado}},\ }\bibfield  {title} {\bibinfo {title} {Topological phase diagrams of exactly solvable non-{Hermitian} interacting {Kitaev} chains},\ }\href {https://doi.org/10.1103/PhysRevResearch.5.L022046} {\bibfield  {journal} {\bibinfo  {journal} {Phys. Rev. Res.}\ }\textbf {\bibinfo {volume} {5}},\ \bibinfo {pages} {L022046} (\bibinfo {year} {2023})}\BibitemShut {NoStop}%
\bibitem [{\citenamefont {Diehl}\ \emph {et~al.}(2011)\citenamefont {Diehl}, \citenamefont {Rico}, \citenamefont {Baranov},\ and\ \citenamefont {Zoller}}]{Zoller2011}%
  \BibitemOpen
  \bibfield  {author} {\bibinfo {author} {\bibfnamefont {S.}~\bibnamefont {Diehl}}, \bibinfo {author} {\bibfnamefont {E.}~\bibnamefont {Rico}}, \bibinfo {author} {\bibfnamefont {M.~A.}\ \bibnamefont {Baranov}},\ and\ \bibinfo {author} {\bibfnamefont {P.}~\bibnamefont {Zoller}},\ }\bibfield  {title} {\bibinfo {title} {Topology by dissipation in atomic quantum wires},\ }\href {https://doi.org/10.1038/nphys2106} {\bibfield  {journal} {\bibinfo  {journal} {Nat. Phys.}\ }\textbf {\bibinfo {volume} {7}},\ \bibinfo {pages} {971} (\bibinfo {year} {2011})}\BibitemShut {NoStop}%
\bibitem [{\citenamefont {Hu}\ \emph {et~al.}(2015)\citenamefont {Hu}, \citenamefont {Cai}, \citenamefont {Baranov},\ and\ \citenamefont {Zoller}}]{Zoller2015}%
  \BibitemOpen
  \bibfield  {author} {\bibinfo {author} {\bibfnamefont {Y.}~\bibnamefont {Hu}}, \bibinfo {author} {\bibfnamefont {Z.}~\bibnamefont {Cai}}, \bibinfo {author} {\bibfnamefont {M.~A.}\ \bibnamefont {Baranov}},\ and\ \bibinfo {author} {\bibfnamefont {P.}~\bibnamefont {Zoller}},\ }\bibfield  {title} {\bibinfo {title} {{Majorana} fermions in noisy {Kitaev} wires},\ }\href {https://doi.org/10.1103/PhysRevB.92.165118} {\bibfield  {journal} {\bibinfo  {journal} {Phys. Rev. B}\ }\textbf {\bibinfo {volume} {92}},\ \bibinfo {pages} {165118} (\bibinfo {year} {2015})}\BibitemShut {NoStop}%
\bibitem [{\citenamefont {Pedrocchi}\ and\ \citenamefont {DiVincenzo}(2015)}]{Divincenzo2015}%
  \BibitemOpen
  \bibfield  {author} {\bibinfo {author} {\bibfnamefont {F.~L.}\ \bibnamefont {Pedrocchi}}\ and\ \bibinfo {author} {\bibfnamefont {D.~P.}\ \bibnamefont {DiVincenzo}},\ }\bibfield  {title} {\bibinfo {title} {{Majorana} braiding with thermal noise},\ }\href {https://doi.org/10.1103/PhysRevLett.115.120402} {\bibfield  {journal} {\bibinfo  {journal} {Phys. Rev. Lett.}\ }\textbf {\bibinfo {volume} {115}},\ \bibinfo {pages} {120402} (\bibinfo {year} {2015})}\BibitemShut {NoStop}%
\bibitem [{\citenamefont {Nayak}\ \emph {et~al.}(2008)\citenamefont {Nayak}, \citenamefont {Simon}, \citenamefont {Stern}, \citenamefont {Freedman},\ and\ \citenamefont {Das~Sarma}}]{DasSarma2008}%
  \BibitemOpen
  \bibfield  {author} {\bibinfo {author} {\bibfnamefont {C.}~\bibnamefont {Nayak}}, \bibinfo {author} {\bibfnamefont {S.~H.}\ \bibnamefont {Simon}}, \bibinfo {author} {\bibfnamefont {A.}~\bibnamefont {Stern}}, \bibinfo {author} {\bibfnamefont {M.}~\bibnamefont {Freedman}},\ and\ \bibinfo {author} {\bibfnamefont {S.}~\bibnamefont {Das~Sarma}},\ }\bibfield  {title} {\bibinfo {title} {Non-abelian anyons and topological quantum computation},\ }\href {https://doi.org/10.1103/RevModPhys.80.1083} {\bibfield  {journal} {\bibinfo  {journal} {Rev. Mod. Phys.}\ }\textbf {\bibinfo {volume} {80}},\ \bibinfo {pages} {1083} (\bibinfo {year} {2008})}\BibitemShut {NoStop}%
\bibitem [{\citenamefont {Pocklington}\ \emph {et~al.}(2023)\citenamefont {Pocklington}, \citenamefont {Wang},\ and\ \citenamefont {Clerk}}]{Clerk2023a}%
  \BibitemOpen
  \bibfield  {author} {\bibinfo {author} {\bibfnamefont {A.}~\bibnamefont {Pocklington}}, \bibinfo {author} {\bibfnamefont {Y.-X.}\ \bibnamefont {Wang}},\ and\ \bibinfo {author} {\bibfnamefont {A.~A.}\ \bibnamefont {Clerk}},\ }\bibfield  {title} {\bibinfo {title} {Dissipative pairing interactions: Quantum instabilities, topological light, and volume-law entanglement},\ }\href {https://doi.org/10.1103/PhysRevLett.130.123602} {\bibfield  {journal} {\bibinfo  {journal} {Phys. Rev. Lett.}\ }\textbf {\bibinfo {volume} {130}},\ \bibinfo {pages} {123602} (\bibinfo {year} {2023})}\BibitemShut {NoStop}%
\bibitem [{\citenamefont {Orr}\ \emph {et~al.}(2023)\citenamefont {Orr}, \citenamefont {Khan}, \citenamefont {Buchholz}, \citenamefont {Kotler},\ and\ \citenamefont {Metelmann}}]{Metelmann2023}%
  \BibitemOpen
  \bibfield  {author} {\bibinfo {author} {\bibfnamefont {L.}~\bibnamefont {Orr}}, \bibinfo {author} {\bibfnamefont {S.~A.}\ \bibnamefont {Khan}}, \bibinfo {author} {\bibfnamefont {N.}~\bibnamefont {Buchholz}}, \bibinfo {author} {\bibfnamefont {S.}~\bibnamefont {Kotler}},\ and\ \bibinfo {author} {\bibfnamefont {A.}~\bibnamefont {Metelmann}},\ }\bibfield  {title} {\bibinfo {title} {High-purity entanglement of hot propagating modes using nonreciprocity},\ }\href {https://doi.org/10.1103/PRXQuantum.4.020344} {\bibfield  {journal} {\bibinfo  {journal} {PRX Quantum}\ }\textbf {\bibinfo {volume} {4}},\ \bibinfo {pages} {020344} (\bibinfo {year} {2023})}\BibitemShut {NoStop}%
\bibitem [{Sup()}]{Supplementary}%
  \BibitemOpen
  \href@noop {} {\bibinfo {title} {See supplementary material}}\BibitemShut {NoStop}%
\bibitem [{Note1()}]{Note1}%
  \BibitemOpen
  \bibinfo {note} {Checking the decay from random initial states this behavior remains qualitatively the same.}\BibitemShut {Stop}%
\bibitem [{\citenamefont {Brighi}\ \emph {et~al.}(2025)\citenamefont {Brighi}, \citenamefont {Ljubotina}, \citenamefont {Roccati},\ and\ \citenamefont {Balducci}}]{Balducci2025}%
  \BibitemOpen
  \bibfield  {author} {\bibinfo {author} {\bibfnamefont {P.}~\bibnamefont {Brighi}}, \bibinfo {author} {\bibfnamefont {M.}~\bibnamefont {Ljubotina}}, \bibinfo {author} {\bibfnamefont {F.}~\bibnamefont {Roccati}},\ and\ \bibinfo {author} {\bibfnamefont {F.}~\bibnamefont {Balducci}},\ }\bibfield  {title} {\bibinfo {title} {Finite steady-state current defies non-{Hermitian} many-body localization},\ }\href {https://doi.org/10.1103/crwj-x7j8} {\bibfield  {journal} {\bibinfo  {journal} {Phys. Rev. Res.}\ }\textbf {\bibinfo {volume} {7}},\ \bibinfo {pages} {L042014} (\bibinfo {year} {2025})}\BibitemShut {NoStop}%
\bibitem [{\citenamefont {Garbe}\ \emph {et~al.}(2024)\citenamefont {Garbe}, \citenamefont {Minoguchi}, \citenamefont {Huber},\ and\ \citenamefont {Rabl}}]{Rabl2024}%
  \BibitemOpen
  \bibfield  {author} {\bibinfo {author} {\bibfnamefont {L.}~\bibnamefont {Garbe}}, \bibinfo {author} {\bibfnamefont {Y.}~\bibnamefont {Minoguchi}}, \bibinfo {author} {\bibfnamefont {J.}~\bibnamefont {Huber}},\ and\ \bibinfo {author} {\bibfnamefont {P.}~\bibnamefont {Rabl}},\ }\bibfield  {title} {\bibinfo {title} {{The bosonic skin effect: Boundary condensation in asymmetric transport}},\ }\href {https://doi.org/10.21468/SciPostPhys.16.1.029} {\bibfield  {journal} {\bibinfo  {journal} {SciPost Phys.}\ }\textbf {\bibinfo {volume} {16}},\ \bibinfo {pages} {029} (\bibinfo {year} {2024})}\BibitemShut {NoStop}%
\bibitem [{\citenamefont {Haga}\ \emph {et~al.}(2021)\citenamefont {Haga}, \citenamefont {Nakagawa}, \citenamefont {Hamazaki},\ and\ \citenamefont {Ueda}}]{Ueda2019c}%
  \BibitemOpen
  \bibfield  {author} {\bibinfo {author} {\bibfnamefont {T.}~\bibnamefont {Haga}}, \bibinfo {author} {\bibfnamefont {M.}~\bibnamefont {Nakagawa}}, \bibinfo {author} {\bibfnamefont {R.}~\bibnamefont {Hamazaki}},\ and\ \bibinfo {author} {\bibfnamefont {M.}~\bibnamefont {Ueda}},\ }\bibfield  {title} {\bibinfo {title} {Liouvillian skin effect: Slowing down of relaxation processes without gap closing},\ }\href {https://doi.org/10.1103/PhysRevLett.127.070402} {\bibfield  {journal} {\bibinfo  {journal} {Phys. Rev. Lett.}\ }\textbf {\bibinfo {volume} {127}},\ \bibinfo {pages} {070402} (\bibinfo {year} {2021})}\BibitemShut {NoStop}%
\bibitem [{\citenamefont {Malz}\ and\ \citenamefont {Nunnenkamp}(2018)}]{Nunnenkamp2018}%
  \BibitemOpen
  \bibfield  {author} {\bibinfo {author} {\bibfnamefont {D.}~\bibnamefont {Malz}}\ and\ \bibinfo {author} {\bibfnamefont {A.}~\bibnamefont {Nunnenkamp}},\ }\bibfield  {title} {\bibinfo {title} {Current rectification in a double quantum dot through fermionic reservoir engineering},\ }\href {https://doi.org/10.1103/PhysRevB.97.165308} {\bibfield  {journal} {\bibinfo  {journal} {Phys. Rev. B}\ }\textbf {\bibinfo {volume} {97}},\ \bibinfo {pages} {165308} (\bibinfo {year} {2018})}\BibitemShut {NoStop}%
\bibitem [{\citenamefont {Leijnse}\ and\ \citenamefont {Flensberg}(2012)}]{Flensberg2012}%
  \BibitemOpen
  \bibfield  {author} {\bibinfo {author} {\bibfnamefont {M.}~\bibnamefont {Leijnse}}\ and\ \bibinfo {author} {\bibfnamefont {K.}~\bibnamefont {Flensberg}},\ }\bibfield  {title} {\bibinfo {title} {Parity qubits and poor man's {Majorana} bound states in double quantum dots},\ }\href {https://doi.org/10.1103/PhysRevB.86.134528} {\bibfield  {journal} {\bibinfo  {journal} {Phys. Rev. B}\ }\textbf {\bibinfo {volume} {86}},\ \bibinfo {pages} {134528} (\bibinfo {year} {2012})}\BibitemShut {NoStop}%
\bibitem [{\citenamefont {ten Haaf}\ \emph {et~al.}(2024)\citenamefont {ten Haaf}, \citenamefont {Wang}, \citenamefont {Bozkurt}, \citenamefont {Liu}, \citenamefont {Kulesh}, \citenamefont {Kim}, \citenamefont {Xiao}, \citenamefont {Thomas}, \citenamefont {Manfra}, \citenamefont {Dvir}, \citenamefont {Wimmer},\ and\ \citenamefont {Goswami}}]{Goswami2024}%
  \BibitemOpen
  \bibfield  {author} {\bibinfo {author} {\bibfnamefont {S.~L.~D.}\ \bibnamefont {ten Haaf}}, \bibinfo {author} {\bibfnamefont {Q.}~\bibnamefont {Wang}}, \bibinfo {author} {\bibfnamefont {A.~M.}\ \bibnamefont {Bozkurt}}, \bibinfo {author} {\bibfnamefont {C.-X.}\ \bibnamefont {Liu}}, \bibinfo {author} {\bibfnamefont {I.}~\bibnamefont {Kulesh}}, \bibinfo {author} {\bibfnamefont {P.}~\bibnamefont {Kim}}, \bibinfo {author} {\bibfnamefont {D.}~\bibnamefont {Xiao}}, \bibinfo {author} {\bibfnamefont {C.}~\bibnamefont {Thomas}}, \bibinfo {author} {\bibfnamefont {M.~J.}\ \bibnamefont {Manfra}}, \bibinfo {author} {\bibfnamefont {T.}~\bibnamefont {Dvir}}, \bibinfo {author} {\bibfnamefont {M.}~\bibnamefont {Wimmer}},\ and\ \bibinfo {author} {\bibfnamefont {S.}~\bibnamefont {Goswami}},\ }\bibfield  {title} {\bibinfo {title} {A two-site {Kitaev} chain in a two-dimensional electron gas},\ }\href {https://doi.org/10.1038/s41586-024-07434-9} {\bibfield  {journal} {\bibinfo  {journal} {Nature}\ }\textbf {\bibinfo {volume}
  {630}},\ \bibinfo {pages} {329} (\bibinfo {year} {2024})}\BibitemShut {NoStop}%
\bibitem [{\citenamefont {Bordin}\ \emph {et~al.}(2024)\citenamefont {Bordin}, \citenamefont {Li}, \citenamefont {van Driel}, \citenamefont {Wolff}, \citenamefont {Wang}, \citenamefont {ten Haaf}, \citenamefont {Wang}, \citenamefont {van Loo}, \citenamefont {Kouwenhoven},\ and\ \citenamefont {Dvir}}]{Dvir2024}%
  \BibitemOpen
  \bibfield  {author} {\bibinfo {author} {\bibfnamefont {A.}~\bibnamefont {Bordin}}, \bibinfo {author} {\bibfnamefont {X.}~\bibnamefont {Li}}, \bibinfo {author} {\bibfnamefont {D.}~\bibnamefont {van Driel}}, \bibinfo {author} {\bibfnamefont {J.~C.}\ \bibnamefont {Wolff}}, \bibinfo {author} {\bibfnamefont {Q.}~\bibnamefont {Wang}}, \bibinfo {author} {\bibfnamefont {S.~L.~D.}\ \bibnamefont {ten Haaf}}, \bibinfo {author} {\bibfnamefont {G.}~\bibnamefont {Wang}}, \bibinfo {author} {\bibfnamefont {N.}~\bibnamefont {van Loo}}, \bibinfo {author} {\bibfnamefont {L.~P.}\ \bibnamefont {Kouwenhoven}},\ and\ \bibinfo {author} {\bibfnamefont {T.}~\bibnamefont {Dvir}},\ }\bibfield  {title} {\bibinfo {title} {Crossed {Andreev} reflection and elastic cotunneling in three quantum dots coupled by superconductors},\ }\href {https://doi.org/10.1103/PhysRevLett.132.056602} {\bibfield  {journal} {\bibinfo  {journal} {Phys. Rev. Lett.}\ }\textbf {\bibinfo {volume} {132}},\ \bibinfo {pages} {056602} (\bibinfo {year}
  {2024})}\BibitemShut {NoStop}%
\bibitem [{\citenamefont {Zatelli}\ \emph {et~al.}(2024)\citenamefont {Zatelli}, \citenamefont {van Driel}, \citenamefont {Xu}, \citenamefont {Wang}, \citenamefont {Liu}, \citenamefont {Bordin}, \citenamefont {Roovers}, \citenamefont {Mazur}, \citenamefont {van Loo}, \citenamefont {Wolff}, \citenamefont {Bozkurt}, \citenamefont {Badawy}, \citenamefont {Gazibegovic}, \citenamefont {Bakkers}, \citenamefont {Wimmer}, \citenamefont {Kouwenhoven},\ and\ \citenamefont {Dvir}}]{Dvir2024a}%
  \BibitemOpen
  \bibfield  {author} {\bibinfo {author} {\bibfnamefont {F.}~\bibnamefont {Zatelli}}, \bibinfo {author} {\bibfnamefont {D.}~\bibnamefont {van Driel}}, \bibinfo {author} {\bibfnamefont {D.}~\bibnamefont {Xu}}, \bibinfo {author} {\bibfnamefont {G.}~\bibnamefont {Wang}}, \bibinfo {author} {\bibfnamefont {C.-X.}\ \bibnamefont {Liu}}, \bibinfo {author} {\bibfnamefont {A.}~\bibnamefont {Bordin}}, \bibinfo {author} {\bibfnamefont {B.}~\bibnamefont {Roovers}}, \bibinfo {author} {\bibfnamefont {G.~P.}\ \bibnamefont {Mazur}}, \bibinfo {author} {\bibfnamefont {N.}~\bibnamefont {van Loo}}, \bibinfo {author} {\bibfnamefont {J.~C.}\ \bibnamefont {Wolff}}, \bibinfo {author} {\bibfnamefont {A.~M.}\ \bibnamefont {Bozkurt}}, \bibinfo {author} {\bibfnamefont {G.}~\bibnamefont {Badawy}}, \bibinfo {author} {\bibfnamefont {S.}~\bibnamefont {Gazibegovic}}, \bibinfo {author} {\bibfnamefont {E.~P.~A.~M.}\ \bibnamefont {Bakkers}}, \bibinfo {author} {\bibfnamefont {M.}~\bibnamefont {Wimmer}}, \bibinfo {author} {\bibfnamefont {L.~P.}\
  \bibnamefont {Kouwenhoven}},\ and\ \bibinfo {author} {\bibfnamefont {T.}~\bibnamefont {Dvir}},\ }\bibfield  {title} {\bibinfo {title} {Robust poor man’s {Majorana} zero modes using {Yu-Shiba-Rusinov} states},\ }\href {https://doi.org/10.1038/s41467-024-52066-2} {\bibfield  {journal} {\bibinfo  {journal} {Nature Communications}\ }\textbf {\bibinfo {volume} {15}},\ \bibinfo {pages} {7933} (\bibinfo {year} {2024})}\BibitemShut {NoStop}%
\bibitem [{\citenamefont {ten Haaf}\ \emph {et~al.}(2025)\citenamefont {ten Haaf}, \citenamefont {Zhang}, \citenamefont {Wang}, \citenamefont {Bordin}, \citenamefont {Liu}, \citenamefont {Kulesh}, \citenamefont {Sietses}, \citenamefont {Prosko}, \citenamefont {Xiao}, \citenamefont {Thomas}, \citenamefont {Manfra}, \citenamefont {Wimmer},\ and\ \citenamefont {Goswami}}]{Goswami2025}%
  \BibitemOpen
  \bibfield  {author} {\bibinfo {author} {\bibfnamefont {S.~L.~D.}\ \bibnamefont {ten Haaf}}, \bibinfo {author} {\bibfnamefont {Y.}~\bibnamefont {Zhang}}, \bibinfo {author} {\bibfnamefont {Q.}~\bibnamefont {Wang}}, \bibinfo {author} {\bibfnamefont {A.}~\bibnamefont {Bordin}}, \bibinfo {author} {\bibfnamefont {C.-X.}\ \bibnamefont {Liu}}, \bibinfo {author} {\bibfnamefont {I.}~\bibnamefont {Kulesh}}, \bibinfo {author} {\bibfnamefont {V.~P.~M.}\ \bibnamefont {Sietses}}, \bibinfo {author} {\bibfnamefont {C.~G.}\ \bibnamefont {Prosko}}, \bibinfo {author} {\bibfnamefont {D.}~\bibnamefont {Xiao}}, \bibinfo {author} {\bibfnamefont {C.}~\bibnamefont {Thomas}}, \bibinfo {author} {\bibfnamefont {M.~J.}\ \bibnamefont {Manfra}}, \bibinfo {author} {\bibfnamefont {M.}~\bibnamefont {Wimmer}},\ and\ \bibinfo {author} {\bibfnamefont {S.}~\bibnamefont {Goswami}},\ }\bibfield  {title} {\bibinfo {title} {Observation of edge and bulk states in a three-site {Kitaev} chain},\ }\href {https://doi.org/10.1038/s41586-025-08892-5}
  {\bibfield  {journal} {\bibinfo  {journal} {Nature}\ }\textbf {\bibinfo {volume} {641}},\ \bibinfo {pages} {890} (\bibinfo {year} {2025})}\BibitemShut {NoStop}%
\bibitem [{\citenamefont {Kawabata}\ \emph {et~al.}(2023)\citenamefont {Kawabata}, \citenamefont {Numasawa},\ and\ \citenamefont {Ryu}}]{Kawabata2023}%
  \BibitemOpen
  \bibfield  {author} {\bibinfo {author} {\bibfnamefont {K.}~\bibnamefont {Kawabata}}, \bibinfo {author} {\bibfnamefont {T.}~\bibnamefont {Numasawa}},\ and\ \bibinfo {author} {\bibfnamefont {S.}~\bibnamefont {Ryu}},\ }\bibfield  {title} {\bibinfo {title} {Entanglement phase transition induced by the non-{Hermitian} skin effect},\ }\href {https://doi.org/10.1103/PhysRevX.13.021007} {\bibfield  {journal} {\bibinfo  {journal} {Phys. Rev. X}\ }\textbf {\bibinfo {volume} {13}},\ \bibinfo {pages} {021007} (\bibinfo {year} {2023})}\BibitemShut {NoStop}%
\end{thebibliography}
\end{document}